\input phyzzx


  \def\tr{{\hbox{\rm Tr}}}

\tolerance=500000 \overfullrule=0pt 
\def\np{Nucl. Phys.}
\def\pl{Phys. Lett.}

  \def\cmp{Comm. Math. Phys.}
 \def\mpl{Mod. Phys. Lett.}   \def\am{Ann. of Math.}  \def\topo{Topology}   \def\jmp{J. Math. Phys.} \def\jgp{J.
Geom. Phys.} \def\jdg{J. Diff. Geom.} \def\mich{Mich. Math. J.}

\def\mrl{Math. Research Lett.}
\def\dmj{Duke Math. J.}

     \def\tr{{\hbox{\rm Tr}}}

\tolerance=500000 \overfullrule=0pt
 
\pubnum={US-FT/12-96 \cr CERN-TH/96-78 \cr hep-th/9603169}
\date={March, 1996} 
\pubtype={}

\titlepage \title{TWISTED $N=2$ SUPERSYMMETRY WITH CENTRAL CHARGE
AND EQUIVARIANT COHOMOLOGY}
\author{J.M.F. Labastida$^{a}$ and  M. Mari\~no$^{a,b}$}
\address{$^{a}$ Departamento de F\'\i sica de Part\'\i culas\break
Universidade de Santiago\break E-15706 Santiago de Compostela,
Spain\break\break $^{b}$ Theory Division, CERN\break CH-1211 Geneva 23,
Switzerland} 

\abstract{We present an equivariant extension of the Thom form 
with respect to a vector field action, in the framework of the Mathai-Quillen 
formalism. The associated Topological Quantum Field Theories correspond 
to twisted $N=2$ supersymmetric theories with a central charge. We analyze in 
detail two different cases: topological sigma models and non-abelian monopoles 
on four-manifolds. }

\endpage  
\pagenumber=1 
\REF\tqft{E. Witten, ``Topological quantum field
theory"\journal\cmp&117(88)353}

\REF\topmodel{E. Witten, ``Topological
sigma models"\journal\cmp&118(88)411}

\REF\mq{V. Mathai and D. Quillen, ``Superconnections, Thom classes, and
equivariant  differential forms"\journal\topo&25(86) 85}

\REF\aj{M.F. Atiyah and L. Jeffrey, ``Topological lagrangians and
cohomology"\journal\jgp&7(90)119}

\REF\donfirst{S.K.
Donaldson, ``An application of gauge theory to four
 dimensional topology"\journal\jdg&18(83)279} 
\REF\don{S.K.
Donaldson, ``Polynomial invariants for smooth
 four-manifolds"\journal\topo&29(90)257} 
\REF\donkron{S.K. Donaldson and P.B.
Kronheimer,  {\it The Geometry of Four-Manifolds}, Oxford Mathematical
Monographs, 1990}
  
\REF\am{P. Aspinwall and D. Morrison,  ``Topological field theory and
rational curves"\journal\cmp&151(93)245}
 
\REF\wu{S. Wu, ``On the
Mathai-Quillen formalism of topological sigma models", hep-th/9406103}
 
\REF\cmrlect{S.
Cordes, G. Moore and S. Rangoolam, Proceedings of the 1994 Les Houches
 Summer School, hep-th/9411210}

\REF\blau{M. Blau and G. Thompson, ``Localization and diagonalization: A
review of functional  integral techniques for low-dimensional  gauge
theories and topological field theories"\journal\jmp&36(95)2192}

\REF\ab{M. Atiyah and R. Bott, ``The moment map and equivariant
cohomology"\journal\topo&23 (84)1} 
\REF\bone{R. Bott, ``Vector fields and
characteristic numbers"\journal\mich&14(67)231} 
\REF\btwo{R. Bott, ``A
residue formula for holomorphic vector fields"\journal\jdg&4(67)311}
\REF\berlin{N. Berline et M. Vergne, ``Zeros d'un champ de vecteurs et
classes characteristiques equivariantes"\journal\dmj&50(83)539}

\REF\llla{J.M.F. Labastida and P.M.
Llatas, ``Potentials for  topological sigma models"\journal\pl&B271(91)101}

\REF\lm{J.M.F. Labastida and M. Mari\~no,  ``Non-abelian monopoles on
four-manifolds"\journal\np&B448(95)373}

\REF\korea{S. Hyun, J. Park and J.S. Park, ``$N=2$ supersymmetric QCD and
four-manifolds; (I) the Donaldson and the Seiberg-Witten invariants",
hep-th/9508162}

\REF\pids{V.
Pidstrigach and A. Tyurin, ``Localisation of the Donaldson's invariants
along  Seiberg-Witten classes", preprint, dg-ga/9507004}
 
\REF\ag{L. Alvarez-Gaum\'e and D.Z. Freedman, ``Potentials for
 the supersymmetric nonlinear sigma model"\journal\cmp&91(83)87}

\REF\ws{N. Seiberg and E. Witten,
``Electric-magnetic duality, monopole condensation, and confinement in $N=2$
$SU(2)$ Yang-Mills theory"\journal\np&B426(94)19; ``Monopole condensation,
duality and chiral symmetry breaking in $N=2$ QCD"\journal\np&B431(94)484 }

\REF\mfm{E. Witten, ``Monopoles and four-manifolds"\journal\mrl&1(94)769}

\REF\ansdos{D. Anselmi and P. Fr\`e, ``Gauged hyperinstantons and monopole 
equations"\journal\pl&B347(95)247}
  
\REF\ot{C.
Okonek and A. Teleman, ``The coupled Seiberg-Witten equations, vortices,
and  moduli spaces of stable pairs", to appear in Int. J. Math.;
``Quaternionic monopoles",  alg-geom/9505209}
  
\REF\tqcd{S. Hyun, J. Park and
J.S. Park, ``Topological QCD"\journal\np&B453(95)199} 
 
\REF\oscar{S. Bradlow and O.
Garc\'\i a Prada, ``Non-abelian monopoles and vortices",  preprint,
alg-geom/9602010}
 
\REF\alab{M. Alvarez and J.M.F. Labastida, ``Breaking of
topological symmetry"\journal\pl&B315(93)251}
   
\REF\alabas{M. Alvarez and
J.M.F. Labastida, ``Topological matter in four
dimensions"\journal\np&B437(95)356}

 \REF\ans{D. Anselmi and P.
Fr\`e, ``Twisted $N=2$ supergravity as topological 
gravity in four dimensions"\journal\np&B392(93)401, 
``Topological twist in four dimensions, 
$R$-duality and hyperinstantons"\journal\np&B404(93)288, ``Topological 
sigma models in 
four dimensions and triholomorphic maps"\journal\np&B416(94)255}

 \REF\wjmp{E. Witten, ``Supersymmetric Yang-Mills theory on a
four-manifold"\journal\jmp&35(94)5101}
 
\REF\lmdos{J.M.F. Labastida and M.
Mari\~no,  ``Polynomial invariants for $SU(2)$
monopoles"\journal\np&B456(95)633}

\REF\zumino{B. Zumino, 
``Supersymmetry and K\"ahler manifolds"\journal\pl&87B(79)203}
 
\REF\luisf{L. Alvarez-Gaum\'e and D. Freedman, ``Geometric 
structure 
and ultraviolet finiteness in the supersymmetric 
sigma model"\journal\cmp&80(81)443}
 
\REF\martin{S.J. Gates, C.M. Hull and M. Rocek, ``Twisted 
multiplets and new 
supersymmetric nonlinear sigma models"\journal\np&B248(84)157}

\REF\egu{T. Eguchi and S.K. Yang, ``$N=2$ superconformal models as 
topological field theories"\journal\mpl&A5(90)1693}

\REF\tmintwo{J.M.F. Labastida and P.M. Llatas, ``Topological 
matter in two dimensions"\journal\np&B379(92)220}

\REF\mima{E. Witten, `Mirror manifolds and topological field theory', {\it in}
Essays on mirror manifolds, ed. S.-T. Yau (International Press, Hong Kong,
1992)}
\REF\vafa{C. Vafa, ``Topological Landau-Ginzburg models"\journal\mpl&A6(91)337}
 
\REF\mart{A. Karlhede and M. Ro\v cek, ``Topological Quantum 
Field Theory and $N=2$ conformal supergravity"\journal\pl&B212(88)51}   

\REF\wv{C. Vafa and E.
Witten \journal\np&B431(94)3 } 

\chapter{Introduction}

It is by now a well-known fact that many $N=2$ supersymmetric theories can 
be reformulated through a ``twisting" of the supersymmetry algebra in order 
to construct 
Topological Quantum Field Theories. The classical examples of this procedure 
are the Donaldson-Witten theory [\tqft] and the topological sigma 
model 
[\topmodel], which arise by twisting the $N=2$ supersymmetric Yang-Mills theory 
and the $N=2$ supersymmetric sigma model, respectively. The most 
useful approach to understand the geometry involved in this kind of 
models is 
perhaps the one based on the Mathai-Quillen formalism [\mq]. It was 
shown in 
[\aj] that the topological lagrangian appearing in 
Donaldson-Witten theory 
can be considered as the Euler class of a certain 
infinite-dimensional bundle 
over the space of Yang-Mills connections. The Euler class is obtained as the 
pullback of the Thom class of the bundle by means 
of a section whose zero 
locus is precisely the moduli space of anti-self-dual instantons of 
Donaldson theory [\donfirst, \don, \donkron]. 
The representative of the 
Thom class that appears in Donaldson-Witten theory is precisely the one 
appearing in [\mq]. Subsequently it was shown that
the same construction holds in the case of 
topological sigma models [\am, \wu]. A review of these
 developments can be found in [\cmrlect, \blau]. 
In the same way, one can use the Mathai-Quillen formalism to construct 
Topological 
Quantum Field Theories starting from a moduli problem formulated in 
a purely geometrical setting. 

However there are some twisted $N=2$ supersymmetric theories which do not 
have a clear formulation in the Mathai-Quillen framework, and therefore their 
geometrical structure is not very well understood. One should then look 
for generalizations of this formalism to take into account the rich 
topological structures hidden in the supersymmetry algebra. The purpose 
of this paper is precisely to obtain an equivariant extension of the Thom 
class of a 
bundle with respect to a vector field action, in the Mathai-Quillen setting. 
This construction can be regarded as a generalization of the equivariant 
extensions of the curvature considered in [\ab, \bone, \btwo, \berlin]. 
Apart form its mathematical interest, it turns out that the Topological 
Quantum Field Theories constructed with this extension 
correspond to twisted $N=2$ 
supersymmetric theories with a central charge. We will consider
in detail two different applications of our construction. The first 
one will be a 
topological sigma model with a vector field action on the target space. The 
resulting theory corresponds to the twisted $N=2$ supersymmetric sigma model 
with potentials constructed in [\llla]. Our second example will be non-abelian 
monopoles on four manifolds [\lm], 
where the vector field action is now given by a 
$U(1)$ symmetry acting on the monopole fields. 
The topological lagrangian that one 
obtains in this way can be regarded as a topological Yang-Mills theory 
coupled to  twisted massive hypermultiplets. This twisted model was 
also considered in [\korea], where the relation to equivariant 
cohomology 
was pointed out. 

These two examples are very interesting from the topological point of view. 
The first one gives the natural framework to consider equivariant quantum 
cohomology of almost-hermitean manifolds with a vector field action. The 
four-dimensional example gives a very explicit connection between $N=2$ 
quantum field theories and the strategy proposed by Pidstrigach and Tyurin 
[\pids] 
to prove the equivalence between Donaldson and Seiberg-Witten invariants 
using non-abelian monopole equations. 

The organization of this paper is as follows. In section 2 we review some 
results on equivariant cohomology and on the construction of equivariant 
extensions of the curvature. In section 3 we present the equivariant 
extension of the Thom form in the Mathai-Quillen formalism. We consider 
different geometrical situations which roughly correspond to the Weil 
or Cartan representatives of the usual Mathai-Quillen form. In section 4 we 
apply the previous results to topological sigma models and non-abelian 
monopoles on four manifolds, from a purely geometrical point of view. 
In section 5 we consider the twisting of $N=2$ supersymmetry with a central 
charge and we relate it to the equivariant cohomology associated to a vector 
field action. We also rederive the two models of section 4 by twisting the
$N=2$ supersymmetric sigma model with potentials and the $N=2$ supersymmetric 
Yang-Mills theory coupled to massive matter hypermultiplets. Finally, 
in section 6 we state our final remarks and conclusions, and some prospects 
for future work.

\endpage
    
\chapter{Equivariant cohomology and equivariant curvature}

\section{Equivariant cohomology}

In this paper we will use the Cartan model for equivariant cohomology, and
here we will review some basic definitions. For a detailed account of
equivariant cohomology, see [\ab, \mq,\cmrlect].

Let $X$ be a vector field acting on a manifold $M$. Recall that every
vector field is associated to a locally defined one-parameter group of
transformations of $M$, $\phi: I\times M \rightarrow  M$, with $I \subset
R$ being an open interval containing $t=0$.  If we put $\phi_m(t)=
\phi_t(m)=\phi(t,m)$,  the vector field corresponding to $\phi$ is given by:
$$
X(m)=\phi_{m * 0} \Big({d \over dt}\Big)_{t=0},
\eqn\vector
$$ 
where $*$ denotes as usual the differential map between tangent spaces. A
particular case of this correspondence is a circle ($U(1)$) action on $M$
with generator $X$.
 
Let ${\cal L}(X)$ be the Lie derivative with respect to the vector field
$X$, and let $\Omega^{*}(M)$ be the complex of differential forms on $M$.
We denote by $\Omega^{*}_{X}(M)$ the kernel of  ${\cal L}(X)$ in
$\Omega^{*}(M)$. We consider now the polynomial ring generated by a
generator $u$ of degree $2$ over  $\Omega^{*}(M)$, denoted by 
$\Omega^{*}(M)[u]$. On this ring we define the equivariant exterior
derivative as follows: 
$$
d_{X} \omega =d\omega - u \iota (X) \omega,\,\,\,\,\,\ \omega \in
\Omega^{*}[u], \eqn\ext
$$
where $\iota (X)$ denotes the usual inner product with the vector field $X$. 
Notice that
$$
d_X^2 =- u{\cal L}(X),
\eqn\nilp
$$
and therefore $d_{X}$ is nilpotent on 
 $\Omega^{*}_{X}(M)[u]$. Elements of $\Omega^{*}_{X}[u]$ are called 
 {\it equivariant differential
 forms}. An equivariant differential form $\omega$ verifying $d_X \omega
=0$ is said to be {\it  equivariantly closed}.   Notice that, if $\omega
\in \Omega^{*}(M)[u]$ and $d_X \omega =0$, necessarily $\omega \in 
\Omega^{*}_{X}(M)[u]$ because of \nilp. 

Given a closed invariant differential form, {\it i.e.}, a form $\omega \in
\Omega^{*}_{X}$ with $d\omega=0$,  we don't get an equivariantly closed
differential form unless $\iota (X) \omega =0$. But it might be possible to
find some polynomial $p$ in the ideal generated by $u$ in 
$\Omega^{*}_{X}(M)[u]$ such that the resulting form $\omega '=\omega +p$ is
equivariantly closed.  The form $\omega '$ is called an {\it equivariant
extension} of $\omega$. One of the purposes  of this paper is to find an
equivariant extension of the Thom class of a vector bundle under suitable
conditions, in the framework of the Mathai-Quillen formalism. As the
Mathai-Quillen form involves the curvature of the vector bundle, we need an
explicit expression for the  equivariant extension of the curvature form.
This has been done by Atiyah and Bott [\ab]  following previous results by
Bott in [\bone, \btwo], and by Berline and Vergne in [\berlin]. Here we
will review this construction for  general vector bundles from the point of
view of equivariant cohomology,  and we will proceed in the same way to
obtain the equivariant extension of the  curvature for principal bundles
[\berlin].  Both  results will be needed in the forthcoming subsections.

\section{Equivariant curvature for vector bundles} 

Let $\pi: E \rightarrow
M$ be a real vector bundle.  We suppose that there is a vector field $X$
acting on $M$, and also an ``action" of this field on  $E$ compatible with
the action on $M$. With this we mean [\ab, \btwo] that there is a
differential operator  $\Lambda$ acting on the space of sections of
 $E$, $\Gamma (E)$: 
$$
\Lambda: \Gamma (E) \rightarrow \Gamma (E),
\eqn\ope
$$
that satisfies the derivation property
$$
\Lambda (fs)=(Xf)s+f \Lambda s,\,\,\,\,\ f \in C^{\infty}(M), \,\ s \in
\Gamma (E).   
\eqn\deriv 
$$
We will be particularly interested in the case in which there is a vector
field $X_E$  acting on $E$ in a compatible way with the action of $X$ on
$M$. With this we mean the following:  let $\hat \phi_t$, $\phi_t$ be the
one-parameter flows corresponding to $X_E$, $X$, respectively. Then the 
following conditions are verified:

i) $\pi \hat \phi_t = \phi_t \pi$, {\it i.e}, the one-parameter flows
intertwine with the  projection map of the bundle.

ii) the map $E_m \rightarrow E_{\phi_t m}$ between fibres is a vector space
homomorphism.

Notice that, if $X_E$, $X$ are associated to circle actions on $E$, $M$,
the above conditions  simply state that $E$ is a $G$-bundle over the
$G$-space $M$, with $G=U(1)$. An obvious consequence of (i) is that $X_E$
and $X$ are $\pi$-related: 
$$
\pi_{*}X_E = X.
\eqn\pirel
$$
When there is a vector field $X_E$ acting on $E$ in the above way the
operator $\Lambda$ is naturally defined as: 
$$
(\Lambda s)(m) = \lim _{t \rightarrow 0} {1 \over t}[s(m) -\hat \phi_t
s(\phi_{-t}(m)) ].  
\eqn\opetwo
$$ 
It's easy to see that, because of condition (i) above, $\hat \phi_t
s(\phi_{-t}(m))$ is in fact  a section of $E$, and using (ii) one can check
that the derivation property \deriv\ holds.  We say that the section $s \in
\Gamma (E)$ is {\it invariant} if $\hat \phi_t s(\phi_{-t}(m))= s(m)$, for
all $t\in I$, $m \in M$. This is equivalent to $\Lambda s =0$. If $s$ is an
invariant  section, $X_E$ and $X$ are also $s$-related: 
$$
s_{*}X = X_E.
\eqn\srel
$$

Consider now a connection $D$ on the real vector bundle $E$ of rank $q$. 
We say that $D$ is {\it equivariant} if it
conmutes with the operator $\Lambda$. Let's write this condition with 
respect to a frame field
$\{ s_i\}_{i=1, \cdots, q}$ on an open set $U \subset M$.
We define the matrix-valued function
and one-form on
$U$, $\Lambda^{j}_i$, $\theta^{j}_{i}$, by:
$$
\Lambda s_i =\Lambda^{j}_is_j,\,\,\,\,\ Ds_i=\theta^{j}_{i}s_j.
\eqn\matrices
$$
Of course, $\theta^{j}_{i}$ is the usual connection matrix. Under a change of
frame $s'=sg$,  where $g \in Gl(q,{\bf R})$, we can use the derivation
property of $\Lambda$ to obtain the matrix  with respect to the new local
frame: $$
\Lambda'=g^{-1} \Lambda g +g^{-1}Xg.
\eqn\changeone$$
Imposing $\Lambda D = D\Lambda$ on
the local frame $\{ s_i\}$ one gets 
$$
d\Lambda^{j}_i+\theta^{j}_{k}\Lambda^{k}_i={\cal L} (X) \theta^{j}_{i}+
\Lambda^{j}_k\theta^{k}_{i}.
\eqn\conm$$
The next step to construct the equivariant curvature is to define an operator
$L_{\Lambda}: \Gamma(E) \rightarrow \Gamma(E)$ given by 
$$
L_{\Lambda}s=\Lambda s -\iota (X)Ds, \,\,\,\,\,\  s \in \Gamma(E).
\eqn\equiope$$
The matrix associated to this operator with respect to a local frame on $U$ is 
$$
(L_{\Lambda})^{j}_i=\Lambda^{j}_i-\theta^{j}_{i}(X).
\eqn\opelocal
$$
Using \changeone\ and the usual transformation rule for the connection matrix 
it is easy to check that $(L_{\Lambda})^{j}_i$ is a tensorial matrix of the
adjoint type: under a change of local frame one has 
$$
L_{\Lambda}'=g^{-1}L_{\Lambda}g.
$$
We will compute now the covariant derivative of the matrix $L_{\Lambda}$.
Using \conm\ we get: $$
\eqalign{
DL_{\Lambda} =&dL_{\Lambda}+[\theta, L_{\Lambda}]\cr
=&d\Lambda + [\theta, \Lambda]-({\cal L}(X)-\iota(X) 
d)\theta-[\theta,\iota(X)\theta]\cr
=&\iota(X) (d\theta +\theta \wedge \theta) =\iota(X) K,\cr}
\eqn\cov
$$
where $K$ is the curvature matrix.

We can introduce now the {\it equivariant curvature} $K_X$ for the vector
bundle case, defined as follows: 
$$
K_X=K+uL_{\Lambda}.
\eqn\eqcurv$$
This is not an equivariant differential form, not even a global differential
form on $M$. To achieve this we have to introduce a symmetric invariant
polynomial  with $r$ matrix entries, $P(A_1, \cdots, A_r)$. Consider 
then the following quantity [\bone]:
$$
P_X = P(K_X, \cdots, K_X) = \sum_{i=0}^{r} u^{i} P_{K}^{(i)},
\eqn\eqpol
$$
where
$$
P_{K}^{(i)}= {r \choose i} P(\overbrace{L_{\Lambda}, \cdots,
 L_{\Lambda}}^{i};K, \cdots, K).
$$
Notice that, as $L_{\Lambda}$ is a tensorial matrix of the adjoint type, $P_X$
is a globally defined differential form in $\Omega^{*}[u]$.   Using \cov\ and
the properties of symmetric invariant polynomials it is easy to prove that 
 $$
\iota(X) P_{K}^{(i)}=dP_{K}^{(i+1)},
\eqn\basic
$$
and from this it follows that $P_X$ is 
an equivariantly closed differential form on $M$.

\section{Equivariant curvature for principal bundles}

Let $\pi: P \rightarrow M$ be a principal bundle with group $G$. We suppose
that  we have two vector fields $X_P$, $X$ acting on $P$ and $M$,
respectively. We will require that the one-parameter flow associated to $X_P$,
${\hat \phi_t}$, conmutes with the right action of $G$ on $P$: 
$$
{\hat \phi_t}(pg)=({\hat \phi_t} p)g, \,\,\,\,\ p\in P,\,\ g \in G.
\eqn\pfbcomp
$$
In this case, if $\phi_t$ is the one-parameter flow associated to $X$ on $M$,
we have $\pi {\hat \phi_t}= \phi_t \pi$, and $X$ and $X_P$ are $\pi$-related.
The vector field $X_P$  is in addition right invariant: 
$$
(X_P)_{pg} = (R_g)_{*p}(X_P)_p.
\eqn\rightinv
$$
Let $\theta$ be a connection one-form on $P$, and consider the function with
values in the  Lie algebra of $G$, $\bf g$, given by $\theta (X_P)=\iota(X_P)
\theta$. Using  \rightinv\ and the properties of the connection it is
immediate to see that $\theta(X_P)$ is
a tensorial zero-form of the adjoint type, {\it i.e.}
$$
\theta (X_P)_{pg} = \theta_{pg} ((R_g)_{*p}(X_P)_p)= ({\rm ad} g^{-1}) \theta
(X_P)_p. \eqn\pfbadj$$
Suppose now that the connection one-form verifies:
$$
{\cal L}(X_P) \theta = 0.
\eqn\pfbeq$$
This is the analog of having an equivariant connection for a  vector bundle.
When \pfbeq\ holds we can construct an equivariant curvature for the principal
bundle in a natural way. First, notice that the covariant derivative of
$\theta (X_P)$ is given by: $$
D \theta (X_P) = -\iota (X_P)K,
\eqn\pfbcov
$$
where $K$ is the curvature associated to $\theta$. 
The equivariant curvature is defined as:
$$
K_{X_P}=K-u\theta (X_P).
\eqn\pfbeqcurv
$$
With the help of an invariant symmetric 
polynomial we can construct the form in 
$\Omega^{*}(P)[u]$:
$$
P_{X_P}=P(K_{X_P}, \cdots, K_{X_P}).
\eqn\pfbinv
$$
Taking into account \pfbcov\ we can proceed as in the vector bundle case and
show  that $ P_{X_P}$ is an equivariantly closed differential form on $P$ with
respect to the action of $X_P$.  On the  other hand, because of \pfbadj\ and
the usual arguments in Chern-Weil theory, $P_{X_P}$  descends to a form in
$\Omega^{*}(M)[u]$, ${\overline P_{X_P}}$. Recall that when a form $\omega$ on
$P$ descends to a form ${\bar \omega}$ on $M$, and the vector fields $X_P$ on
$P$ and $X$ on $M$ are $\pi$-related, we have the following identities:
$$
\eqalign{
&{\bar \phi}(V_1, \cdots, V_q)=\phi(X_1, \cdots, X_q),\cr
&(d{\bar \phi})(V_0, \cdots, V_q)=(d \phi) (X_0, \cdots, X_q),\cr
&(\iota(X){\bar \phi})(V_1, \cdots, V_{q-1})=(\iota(X_P)\phi ) 
(X_1, \cdots, X_{q-1}),\cr
&({\cal L}(X){\bar \phi})(V_1, 
\cdots, V_{q})=({\cal L}(X_P)\phi ) (X_1, \cdots, X_{q}),\cr}
\eqn\ident
$$
where the $X_i$ are such that $\pi_{*}X_i=V_i$. It follows from \ident\ that,
if $P_{X_P}$  is equivariantly closed on $P$, then ${\overline P_{X_P}}$ is
equivariantly closed on $M$. We have therefore obtained an appropriate
equivariant extension of the curvature of a principal bundle.

\endpage

\chapter{Equivariant extensions of the Thom form in the Mathai-Quillen
formalism} 

We begin this section with a quick review of the Mathai-Quillen
formalism.  Most of the details will appear in the explicit constructions of
the equivariant extensions of the Thom form for a vector field action, so we
just recall some results. A more complete presentation can be found in [\mq,
\aj, \cmrlect]. 
 
\section{The Mathai-Quillen formalism}

The Mathai-Quillen formalism [\mq] provides an explicit representative of the
Thom form of a vector bundle $E$. Usually this form is introduced in the
following way: consider an oriented vector bundle $\pi : E \rightarrow M$ with
fibre $V={\bf R}^{2m}$, equipped with an inner product $g$ and a compatible
connection $D$.  Let $P$ be the principal $G$-bundle over $M$ such that $E$ is
the associated vector bundle. Then we can consider the $G$-equivariant
cohomology of $V$ in the Weil model, and we introduce the generators $K$ and
$\theta$ for the Weil complex  ${\cal W}(\bf g)$, of degree two and one,
respectively.  As our vector bundle is oriented and has  an inner product, we
can reduce the structural group to $G=SO(2m)$. The universal Thom form $U$ of
Mathai and Quillen is an element in ${\cal W}({\bf g}) \otimes \Omega^{*}(V)$
given by:   
$$
U= (2 \pi)^{-m}{\rm Pf}(K){\rm exp}\{ -x_ix_i-(dx_i +\theta
_{il}x_l)(K^{-1})_{ij} (dx_j + \theta_{jm}x_m)\}, \eqn\Thomform
$$
where 
$x_i$ are orthonormal coordinate functions on $V$, and $dx_i$ are 
their corresponding differentials. This expression 
includes the inverse of $K$, and in fact it
should be properly understood, once the exponential is expanded, as:
$$
{\pi}^{-m}{\rm e}^{ -x_ix_i}\sum_{I} \epsilon(I,I'){\rm Pf}({1 \over 2} K_{I})
(dx +\theta x)^{I'},
\eqn\Uromano
$$
where $I$ denotes a subset with an even number of indices, $I'$ its complement
and $\epsilon(I,I')$ the signature of the corresponding permutation.  The
equivalence of the two representations is easily seen using Berezin
integration. Of course, the expression \Thomform\ is easier to deal with, and
in fact we can check its properties taking $K^{-1}$ as a formal  inverse of
$K$. This is because we can consider \Thomform\ as an element of the  ring of
fractions with ${\rm det}(K)$ in the denominator. Being ${\rm det}(K)$ closed,
we can extend the exterior derivative as an algebraic operator to this
localization [\mq]. We will use later this  principle to check the
equivariantly closed character of our extension. One can also obtain a 
universal Thom form in the Cartan model of the $G$-equivariant cohomology, by 
putting the generator $\theta$ to zero. This gives an alternative
representative which  is useful in topological gauge theories [\aj, \cmrlect].

The form \Thomform\ can be mapped to a  differential form in $\Omega (P \times
V)$ using the Weil homomorphism. This amounts to substituting the algebraic
generators of the Weil complex,  $K$ and $\theta$, by the actual curvature and
connection of the principal bundle $P$. The  resulting form descends to $E$
and gives an explicit representative of the Thom form of $E$, which will be
denoted by $\Phi (E)$. If one  uses the Cartan representative, one must
enforce in addition a horizontal projection.

\section{Equivariant extension of the Thom form: general case}

One of the purposes of this paper is to find an equivariant extension of the
Thom form for  a vector field action, in the framework of the Mathai-Quillen
formalism. If we look at the expressions for  the equivariant extension of the
curvature, \eqcurv\ and \pfbeqcurv, we see that they  involve the contraction
of the connection form with a vector field. It is clear that for  the
algebraic elements in the Weil algebra this operation is not defined, and
therefore we won't work with the universal Thom form, but with the explicit
Thom form as an element of $\Omega^{2m}(E)$. This has also the advantage of
showing explicitly the geometry involved in the Mathai-Quillen formalism,
which is sometimes hidden behind the use of $G$-equivariant cohomology.

Recall that we defined an ``action" of a vector field on a vector bundle $E$
as an operator acting on the space of sections of this bundle, and therefore
not necessarily induced by an action of $X_E$ on $E$. In this case we cannot
consider the complex $\Omega^{*}_{X_E}[u]$. However, given an invariant
section $s$ of this bundle,  we can construct an equivariant extension of the
pullback $s^{*}\Phi (E)$ on $M$. 

In the framework of the Mathai-Quillen formalism we need an inner product on
$E$, $g$, and a compatible  connection $D$ verifying: $$
d(g(s,t))=g(Ds,t)+g(s,Dt),\,\,\,\,\,\ s,t \in \Gamma(E)
\eqn\comp
$$
Once we take into account the action of a vector field $X$ on $M$, and the
compatible  operator on sections $\Lambda$, we need additional assumptions to 
construct our equivariant extensions.  First of all, we assume, as in the
previous subsection, that the connection $D$ is equivariant.  We also assume
that the inner product is invariant with respect to the compatible actions: 
$$
{\cal L}(X)(g(s,t))=g(\Lambda s, t)+ g(s, 
\Lambda t),\,\,\,\,\,\ s,t \in \Gamma(E)
\eqn\cometrica
$$
From \comp\ and \cometrica\ one gets the following 
identity for the operator $L_{\Lambda}$ defined
in \equiope:
$$
g(L_{\Lambda}s,t)+g(s,L_{\Lambda}t)=0,\,\,\,\,\,\ s,t \in \Gamma(E)
\eqn\antiope
$$
We suppose that our bundle $E$ is orientable, 
and therefore we can reduce the structural group
to $SO(2m)$ and consider orthonormal frames 
$\{s_i\}_{i=1, \cdots, 2m}$ such that $g(s_i, s_j)=
\delta_{ij}$. With respect to an orthornormal frame, 
the connection and curvature matrices
are antisymmetric, and because of \antiope\ 
$L_{\Lambda}$ and $K_X$ are antisymmetric too. 

Consider now a trivializing open covering of $M$, $\{U_{\alpha} \}$,  and the
corresponding orthonormal frames $\{ s_i^{\alpha}\}$. Let $s \in \Gamma(E)$ be
an invariant section.  Then, the following form  is an equivariantly closed
differential form on $M$ and is an equivariant extension of the pullback of the
Thom class by $s$: 
$$
 s^{*} \Phi(E)^{\alpha}_{X}= (2 \pi)^{-m}{\rm Pf}(K_X){\rm exp}
\{ -\xi^{\alpha}_i\xi^{\alpha}_i
-(d\xi^{\alpha}_i +\theta ^{\alpha}_{il}
\xi^{\alpha}_l)(K_{X}^{\alpha})^{-1}_{ij}
(d\xi^{\alpha}_j + \theta^{\alpha}_{jm}\xi^{\alpha}_m)\},
\eqn\eqoneThom
$$
where $s=\xi_i s_i^{\alpha}$ is the local expression of $s$ in $U_{\alpha}$,
and $\theta^{\alpha}$ and $K_{X}^{\alpha}$ are respectively the connection and
the  equivariant curvature matrices  (the equivariant curvature is the one
given in \eqcurv). Both are defined with respect to the orthonormal frame $\{
s_i^{\alpha}\}$.  

To prove our statement, we will show first of all that the  $ s^{*}
\Phi(E)^{\alpha}_{X}$ define a global differential form on $M$, {\it i.e.}, we
will consider a change of  trivialization on the intersections $U_{\alpha}
\cap U_{\beta}$.  The transformations of the different functions  appearing
here are: 
$$\eqalign{
&s^{\beta}=s^{\alpha}g_{\alpha \beta}, 
\,\,\,\,\,\ \xi^{\beta}=g^{-1}_{\alpha \beta}\xi^{\alpha},\cr
&\theta^{\beta}=g_{\alpha \beta}^{-1} \theta^{\alpha} g_{\alpha \beta} 
+g_{\alpha \beta}^{-1}dg_{\alpha \beta},\cr
&K_{X}^{\beta} =g_{\alpha \beta}^{-1}K_{X}^{\alpha}g_{\alpha \beta},\cr}
\eqn\trans
$$
where $g_{\alpha \beta}$ are the transition functions and take values in
$SO(2m)$.  To check the invariance of \eqoneThom\ under this transformation, 
notice that ${\rm Pf}(K_X)$ is an invariant symmetric polynomial for 
antisymmetric matrices and therefore the results of section 2 hold. Also
notice that  $d\xi^{\alpha}_i +\theta ^{\alpha}_{il}\xi^{\alpha}_l$ transforms
as a tensorial matrix of the adjoint type  (because it is the local expression
of the covariant derivative $Ds$). It is easily checked that  $ s^{*}
\Phi(E)^{\alpha}_{X}$ equals $ s^{*} \Phi(E)^{\beta}_{X}$ on the intersections
$U_{\alpha} \cap U_{\beta}$,  and therefore the expression \eqoneThom\ defines
a global differential form on $M$.

To prove that this differential form is in the kernel of $d_{X}$ it  is 
enough to do it for the local expression in \eqoneThom, as $d_X$ is a local
operator. Again, by the results of section 2, ${\rm Pf}(K_X)$ is already
equivariantly closed, and we only need to check this property for the exponent
in \eqoneThom. The computation is lengthy but straightforward. Recall that $s$
is an invariant section, and locally this can be written as: 
$$
\Lambda (\xi_i s_i)=X(\xi_i)s_i+\xi_i\Lambda_{ji}s_j=0.
\eqn\lolita
$$
It follows then that
$$
d_{X}d\xi_i=-uX(\xi_i)=-u\Lambda_{ij} \xi_j,
\eqn\deman
$$
and we get the following expression:
$$
\eqalign{ 
&d_{X} \{  \xi_i \xi_i+(d \xi_i +\theta _{il}\xi_l)(K_{X}^{-1})_{ij}
(d\xi_j + \theta_{jm}\xi_m) \} \cr
&=2 d\xi_i \xi_i + [u ( \Lambda_{il}\xi_l -\theta_{il}(X) \xi_l)+
d\theta_{il}\xi_l-\theta_{il}\xi_l]
(K_{X}^{-1})_{ij}(d\xi_j + \theta_{jm}\xi_m)\cr
&-(d\xi_i +\theta _{il}\xi_l)(d_{X}K_{X}^{-1})_{ij} 
(d\xi_j + \theta_{jm}\xi_m)\cr
&-(d\xi_i +\theta _{il}\xi_l)(K_{X}^{-1})_{ij}
[u (\Lambda_{jm}\xi_m -\theta_{jm}(X) \xi_m)+
 d\theta_{jm}\xi_m-\theta_{jm}\xi_m].\cr}
\eqn\calculo
$$
If we add to this $(\theta_{il}+\theta_{li})(d\xi_l 
+\theta _{lp}\xi_p)(K_{X}^{-1})_{ij}
(d\xi_j +\theta _{jm}\xi_m)=0$, and we take into account that
$$
D(K_{X}^{-1})_{ij}=d(K_{X}^{-1})_{ij}+
\theta_{il}(K_{X}^{-1})_{lj}-(K_{X}^{-1})_{il}\theta_{lj},
\eqn\uapiti
$$
then \calculo\ reads:
$$
\eqalign{ 
&2 d\xi_i \xi_i+(K_{X})_{il}\xi_l 
(K_{X}^{-1})_{ij}(d\xi_j + \theta_{jm}\xi_m)\cr
&-(d\xi_i +\theta _{ip}\xi_p)[D(K_{X}^{-1})_{ij}-u
\iota(X) (K_{X}^{-1})_{ij}]
(d\xi_j + \theta_{jm}\xi_m)\cr
&+(d\xi_i +\theta _{ip}\xi_p)(K_{X}^{-1})_{il}(K_{X})_{lm}\xi_m.\cr}
\eqn\marisa
$$
We can compute $DK_{X}^{-1}-u\iota(X) K_{X}^{-1}$ considering $K_{X}^{-1}$
a formal inverse of $K_X$. Notice first that, because 
of the Bianchi identity and \cov,  
we have:
$$
DK_X=u\iota(X) K, \,\,\,\,\ d_{X}K_X=-[\theta, K_X].
\eqn\covx
$$
As $d_X$ extends to the ring of fractions with ${\rm det} K_X$ 
in the denominator (because
${\rm det} K_X$ is $d_X$-closed), we have:
$$
d_{X}K_{X}^{-1}=K_{X}^{-1}[\theta, K_X]K_{X}^{-1}=-[\theta, K_{X}^{-1}],
\eqn\judith
$$
and finally we get:
$$
DK_{X}^{-1}-u\iota(X) K_{X}^{-1}=d_{X}K_{X}^{-1}+[\theta, K_{X}^{-1}]=0
\eqn\clarita
$$
Using \clarita\ and the antisymmetry of the matrices 
$(K_{X})_{ij}$, $\theta_{ij}$, we see that
\marisa\ equals zero. Therefore, \eqoneThom\ is 
in the kernel of $d_X$, and according to \nilp\ 
it is an equivariantly closed differential form. 
It is clear that it is an equivariant extension
of the pullback $s^{*} \Phi(E)$, because if we put $u=0$ 
we recover the pullback of the Mathai-Quillen
form. 

\section{Equivariant extension of the Thom form: vector bundle case}

Now we will consider the case in which we have a vector field 
$X_E$ acting on the vector bundle $E$,
and the action $\Lambda$ is the one induced from it. 
In this case it makes sense to construct an 
equivariant extension of the Thom form with respect to 
the $X_E$ action. Again we will proceed 
locally and we will construct the extension on 
trivializing open sets $U_{\alpha} \times V$.

Let $\pi: E \rightarrow M$ be an orientable real vector bundle of rank $2m$
with  an action of a vector field $X_E$ compatible with an action of $X$ on
$M$ in the sense of  subsection 2.2. On the fibre $V={\bf R}^{2m}$ we choose
an orthonormal basis $\{e_i\}$ with respect to the  standard inner product
$(,)$ on it,  and we denote by $x_i$ the coordinate functions with respect to
this basis. Let $\{U_{\alpha}\}$ be a trivializing open covering of $M$, with
attached diffeomorphisms 
$$
\phi_{\alpha}:U_{\alpha} \times V \rightarrow \pi^{-1} (U_{\alpha}).
\eqn\trivial
$$
If $g$ is the metric on $E$, we can reduce 
the structural group in such a way that 
$g(\phi_{\alpha}(m,v),
\phi_{\alpha}(m,w))=(v,w)$. This also gives an 
orthonormal frame for each $U_{\alpha}$ in the
standard way:
$$
s_i^{\alpha}(m)=\phi_{\alpha}(m,e_i).
\eqn\frame
$$ We want to define a vector field action
${\hat X}_{\alpha}$ on each $U_{\alpha} \times V$ such that
$$
(\phi_{\alpha}^{-1})_{*}(X_E)={\hat X}_{\alpha}.
\eqn\abre
$$
To do this we will define a one-parameter flow ${\hat \phi_t}$ inducing ${\hat
X}_{\alpha}$.  The natural way is to use the conditions of  compatibility of
the vector field actions. On the first factor, $U_{\alpha}$, we use the
restriction of one-parameter flow associated to $X$,  and we take the
appropriate $t$-interval for this map to be well defined. On the second factor
we use the homomorphism between fibres given by the one-parameter flow
associated to $X_E$, $\phi_t^{E}$. Written in a local trivialization, this
homomorphism  means that, if $p \in E_m$, $\phi_t^{E}p \in E_{\phi_t m}$, then
$$
(\pi_2 \phi_{\alpha}^{-1})(\phi_t^{E}p)=\lambda (t, m) 
\pi_2\phi_{\alpha}^{-1}(p),
\eqn\homo
$$
where $\pi_2$ denotes the projection of $\phi_{\alpha}^{-1}$ 
on the second factor and $\lambda$
is an endomorphism of $V$ which depends on $t$, the basepoint 
$m$ and the trivialization. Now 
we can define:
$$
{\hat \phi_t} (m,v) =(\phi_t (m),\lambda (t, m) v), 
\,\,\,\,\ (m,v) \in U_{\alpha} \times V.
\eqn\flujo
$$
Notice that the endomorphism $\lambda$ verifies:
$$
\lambda (s,\phi_t(m)) \lambda (t,m)=\lambda (s+t, m).
\eqn\compositio
$$
From the definition of ${\hat \phi_t}$ it follows that 
$$
\phi_{\alpha}^{-1}\phi_t^{E}= {\hat \phi_t} \phi_{\alpha}^{-1},
\eqn\conm
$$
and this in turn implies \abre. 

The procedure is now similar to the one presented in the preceding 
section. We define the 
following form on $\Omega^{*}(U_{\alpha} \times V)[u]$:
$$
 \Phi(E) ^{\alpha}_{X}= (2 \pi)^{-m}{\rm Pf}(K_X){\rm exp}
\{ -x_i x_i
-(dx_i +\theta_{il}x_l)(K_{X}^{\alpha})^{-1}_{ij}
(dx_j + \theta_{jm}x_m)\},
\eqn\eqdosThom
$$
where $\theta_{ij}$, $(K_{X})_{ij}$ denote respectively the connection and
equivariant  curvature matrices associated to the orthonormal frame defined in
\frame. The index $\alpha$ labeling the trivialization is understood.  We want
to check that \eqdosThom\ defines a global differential form on $E$. First we
will consider the behavior of $\omega^{\alpha}=
 \Phi(E) ^{\alpha}_{X}$ under a change of trivialization. The transition
functions  for the vector bundle are defined as  $g_{\beta
\alpha}=\phi_{\beta}^{-1} \phi_{\alpha}$, restricted as usual to  $\{ x
\}\times V$. The behavior of the connection and curvature matrices under
the change of trivialization is given in \trans, and the gluing  conditions
for the elements in the trivializing open sets are 
$$
(m, v)^{\beta}=(m, g_{\alpha \beta}^{-1} (v))^{\alpha}.
\eqn\gluon
$$
The coordinate functions then transform as 
$x_i \rightarrow (g_{\alpha \beta}^{-1})_{ij}x_j$.
Following the same steps as in the preceding 
section we see that the forms $\omega^{\alpha}$
do not change when we go from the $\alpha$ description to 
the $\beta$ description:
$$
g_{\alpha \beta}^{*} \omega^{\alpha} =\omega^{\beta}.
\eqn\cambia
$$
The forms $\omega^{\alpha}$ define the corresponding forms on 
$\pi^{-1}(U_{\alpha})$ by taking
$(\phi_{\alpha}^{-1})^{*}\omega^{\alpha}$ on these open sets. 
On the intersections we have, because
of \cambia,
$$
(\phi_{\alpha}^{-1})^{*}\omega^{\alpha}=(\phi_{\beta}^{-1})^{*}\omega^{\beta},
\eqn\nocambia
$$
and therefore they define a global differential form on $E$. Now it is clear
that, if the $\omega^{\alpha}$ are in the kernel of $d_{\hat X}$, the
$(\phi_{\alpha}^{-1})^{*} \omega^{\alpha}$ are in the kernel of $d_{X_E}$. 
This is a consequence of the following simple result: if $f: M \rightarrow N$
is a differentiable map, $\omega \in \Omega^{*}(N)$, and $X_M$, $X_N$ are two
vector fields which are $f$-related, then
$$
\iota(X_M) f^{*}\omega = f^{*} \iota(X_N) \omega.
\eqn\simple
$$
Using \simple\ and \abre\ we see that
$$
d_{X_E}(\phi_{\alpha}^{-1})^{*}
\omega^{\alpha}=(\phi_{\alpha}^{-1})^{*}
\big(d-\iota ((\phi_{\alpha}^{-1})_{*}(X_E))\big)\omega^{\alpha}=
(\phi_{\alpha}^{-1})^{*}(d_{\hat X} \omega^{\alpha}).
\eqn\tudobem
$$
To prove that the $\omega^{\alpha}$ are in the kernel of $d_{\hat X}$, notice
that the computation  is very similar to the one presented in the preceding
section. The only new thing we must compute is $d_{\hat X}(dx_i) = -u {\cal
L}({\hat X})x_i$. Using the definition of Lie  derivative and the action of
the one-parameter group associated to ${\hat X}$ and given in  \flujo, we get:
$$
({\cal L}({\hat X})x_i)(m,v)=
-{d \over dt}\lambda_{ij}(-t, m)\Big|_{t=0}x_j(v).
\eqn\lie
$$
The matrix appearing in this expression is not new. To see it, notice that the
matrix representation of the operator $\Lambda$ with respect to the
orthonormal frame \frame\ is given by 
$$
(\Lambda s_i^{\alpha})(m)= 
\lim_{t \rightarrow 0} { s_i^{\alpha}(m)-\phi_t^{E} 
s_i^{\alpha}(\phi_{-t} m) \over t}.
\eqn\opeframe
$$
Using \frame\ and \conm\ we obtain:
$$
\phi_t^{E} 
s_i^{\alpha}(\phi_{-t} m)=\phi_{\alpha} 
{\hat \phi_t} (\phi_{-t}m, e_i)=s_j^{\alpha}(m)
\lambda_{ji}(t,\phi_{-t} m),
\eqn\calculillo
$$
and this gives
$$
(\Lambda s_i^{\alpha})(m)=-s_j^{\alpha}(m)
{d \over dt} \lambda_{ji}(t,\phi_{-t} m) \Big|_{t=0}
\eqn\culillo
$$
Finally, using \compositio\ and comparing \lie\ and \culillo\ we get
$$
({\cal L}({\hat X})x_i)(m,v)=-\Lambda_{ij}(m) x_j(v).
\eqn\final
$$
If we compare this expression to \deman\ we see  that the computation of the
equivariant  exterior derivative of $\omega^{\alpha}$ with  respect to ${\hat
X}$ simply mimicks the one we did in the preceding section. Therefore, the
forms defined in \eqdosThom\ are in the kernel  of $d_{\hat X}$ and the global
differential form $\Phi(E)_{X}$ they induce on $E$ is an equivariantly closed
differential form because of \tudobem. It clearly equivariantly extends  the
Mathai-Quillen expression for the Thom form of the bundle.

Consider now an  invariant section $s \in \Gamma (E)$. Because of \srel\ and
\simple\ it is easy to see that  $s^{*}\Phi(E)_X$ is an equivariantly closed
differential form on the base manifold $M$. Of  course the local expression of
this form coincides with \eqoneThom: the map  $\phi^{\alpha}s: U_{\alpha}
\rightarrow U_{\alpha} \times V$ is given by  
$$
(\phi^{\alpha}s)(m) = (m, \xi_i^{\alpha}(m)e_i),
\eqn\localmap
$$
where we wrote $s=\xi_i^{\alpha}s^{\alpha}_i$. 
As the local expression of $s^{*}\Phi(E)_X$ 
is
$$
(s^{*}\Phi(E)_X)^{\alpha}=(\phi^{\alpha}s)^{*} \Phi(E) ^{\alpha}_{X},
\eqn\maslocal
$$
and from \localmap\ this amounts to substitute $x_i$ 
by $\xi_i^{\alpha}$ in \eqdosThom, we recover precisely
\eqoneThom.

FInally, we will give a field theory expression for $\Phi(E)_X$ using Berezin
integration.  Introduce Grassmann variables $\rho_i$ for the local coordinates
of the fibre. The standard rules of Berezin integration [\mq, \cmrlect] give
the following representative for the local expression \eqdosThom: 
$$
\Phi(E)^{\alpha}_X = \pi^{-m} {\rm e}^{-x_i x_i} \int {\cal D} \rho \,\
{\rm exp} \Big( {1 \over 4} \rho_i K_{ij} \rho_j +
{u \over 4}\rho_i(L_{\Lambda})_{ij}\rho_j
+i(dx_i +\theta_{ij}x_j)\rho_i \Big).
\eqn\campos
$$
With this expression at hand, one can also introduce 
the standard objects in topological field theory,
namely a gauge fermion and a BRST complex. 
Following [\cmrlect], we introduce an auxiliary field
$\pi_i$ with the meaning of a basis of differential 
forms $dx_i$ for the fibre. The BRST
operator is given by the $d_{\hat X}$ cohomology, and therefore we have:
$$
Q\rho_i =\pi_i, \,\,\,\,\,\,\ Q\pi_i=u\Lambda_{ij}\pi_j.
\eqn\brstfibra
$$
On the original fields $x_i$ and the matrix-valued functions on 
$U_{\alpha}$, $\theta_{ij}$, 
$K_{ij}$, $(L_{\Lambda})_{ij}$, $Q$ acts 
again as $d_{\hat X}$. The gauge fermion is the same
than the gauge fermion in the Weil model for the Mathai-Quillen 
formalism [\cmrlect]:
$$
\Psi = -\rho_i (ix_i -{1 \over 4}\theta_{ij} \rho_j +{1 \over 4}\pi_i), 
\eqn\gauge
$$
and it is easily checked that $Q \Psi$ gives, 
after integrating out the auxiliary field $\pi_i$,
the exponent in \campos. This representative will be useful to construct the
equivariant  extension for topological sigma models. Notice that in the
expression \campos\ we can  work with a non-orthonormal metric on $V$ by
introducing the corresponding jacobian in the  integration measure.

\section{Equivariant extension of the Thom form: principal bundle case}

 We
will consider, finally, the case in which the vector bundle $E$ is explicitly
given as an associated vector bundle to a principal bundle $\pi: P \rightarrow
M$, {\it i.e.}, we consider the action of the  structural group $G$ on
$P\times V$ given by $(p,v)g=(pg, g^{-1}v)$, and we form the quotient
$E=P\times V / G$. Notice that $P\times V$ can be considered as a principal 
bundle over $E$. We assume that we have a vector field action on $P\times V$
whose one-parameter flow  $\mu_t$ has the following structure: 
$$
\mu_t (p,v)=(\phi_t^{P}p, \lambda (t,p)v) \,\,\,\,\,\ p \in P, v \in V,
\eqn\masflujo
$$
where $\lambda (t,p)$ is an endomorphism of $V$. 
We also assume that this flow conmutes 
with the $G$-action on $P\times V$:
$$
(\phi_t^{P}p)g=\phi_t^{P}(pg), \,\,\,\,\,\ \lambda 
(t,pg)=g^{-1}\lambda (t,p) g.
\eqn\conmutamas
$$
Because of the above condition, a vector field action on $E$ is induced in the
natural way, and the one-parameter flow $\phi_t^{P}$ gives in turn a vector
field action on $M=P/G$ in the way considered in subsection 2.3, with
one-parameter flow $\phi_t$. In addition, with  these assumptions, the vector
field action on $E$ is compatible with the vector field action on $M$
according to our definition  in subsection 2.2. Condition (i) is immediate,
and to see that condition (ii) holds consider  a trivializing open covering
for $M$, $\{U_{\alpha}\}$, and the corresponding map $\nu_{\alpha}:
\pi^{-1}(U_{\alpha}) \rightarrow G$.  If $m \in U_{\alpha}$, $\phi_t(m) \in
U_{\beta}$,  the map between the fibres $E_m$, $E_{\phi_t m}$ is given by the
homomorphism 
$$
\nu_{\beta}(\phi_t^{P}p) \lambda (t,p) \nu_{\alpha}(p)^{-1}, \eqn\endo$$ where
$p \in \pi^{-1}(m)$. Using \conmutamas\ it is easy to see that \endo\  only
depends on the basepoint $m$ and $t$. The vector fields on $P$, $E$ and $M$
will  be denoted, respectively,  by $X_{P}$, $X_E$ and $X$.  Our last
assumption is that there is an inner product $(,)$ on $V$ preserved  by both
the action  of $G$ and the endomorphisms $\lambda(t,p)$. As usual, this means
that the matrix  
$$
\Lambda_{ij}(p)=\lim_{t \rightarrow 0}{1 \over t} 
[\delta_{ij}-\lambda (t, p)_{ij}]
\eqn\openuevo
$$
is antisymmetric, where the components are taken with
 respect to an orthonormal basis 
${e_i}$
of $V$. If we regard $P\times V$ as a principal bundle, the second 
condition in \conmutamas\
imply that $\Lambda$ is a tensorial matrix of the
adjoint type. 

We will be particularly interested in the case in which $\lambda (t,p)$
doesn't depend on $p$. In this case we have that $\Lambda$ is a constant
matrix  conmuting with all the $g \in G$ (and then with all the elements in
the Lie algebra $\bf g$).  This happens, for instance, if $G=U(m) \subset
SO(2m)$ and $\Lambda$ has the structure: $$
\Lambda=\left( \matrix{ 0&{1}&{\ldots}& & \cr
                      {-1}&0&{\ldots}& & \cr
                      {\vdots}&{\vdots}&{\ddots}\cr
                        & &{\ldots}&0&{1} \cr
                        & &{\ldots}&{-1}&0 \cr} \right).                
\eqn\matriz
$$
This is in fact the situation we will find 
in the application of our formalism to 
non-abelian monopoles on four-manifolds.

Let $\theta$ and $K$ be respectively the connection and curvature of $P$.
Assume now, as 
 in subsection 2.3, that ${\cal L}(X_P)\theta=0$, 
 and that $\Lambda$ is a constant matrix conmuting with
 all the $A \in {\bf g}$. Then $D\Lambda =0$. 
 We want to construct an equivariantly closed differential
 form on $P \times V$ with respect to the vector field action ${\hat X}=(X_P,
X_V)$, where $X_V$ 
 is associated to the flow $\lambda(t)$. First of all we define an equivariant
curvature on 
 $P \times V$:
 $$
 K_X=K+u(\Lambda-\theta(X_{P})).
 \eqn\claudia
 $$
Notice that $\Lambda -\theta(X_{P})$ is a tensorial matrix of 
the adjoint type, and
if $P(A_1, \cdots, A_r)$ is an invariant symmetric 
polynomial for the adjoint action of ${\bf
 g}$, then we can go through the arguments of subsection 2.3 to show that 
 $P( K_X, \cdots, K_X)$ defines an equivariantly 
 closed differential form on $P \times V$. 
 The construction of the equivariant extension 
 of the Thom class is very similar to the ones
 we have done before, but now we define a form on 
 $P\times V$ and we will show that it descends
 to $E$. Consider then the following element in $\Omega^{*}(P \times V)[u]$:
 $$
 \Phi (P\times V)=(2 \pi)^{-m}{\rm Pf}(K_X){\rm exp}
\{ -x_i x_i
-(dx_i +\theta_{il}x_l)(K_{X}^{\alpha})^{-1}_{ij}
(dx_j + \theta_{jm}x_m)\},
\eqn\eqtresThom
$$
where $x_i$ are, as before, orthonormal coordinates 
on the fibre $V$. First we will check that
the above form descends to $E$. For this we must 
check that it is right invariant and that it vanishes
on vertical fields.  
The first property is easily checked using the expressions:
$$
(R_g^{*} x_i)(v)=x_i(g^{-1}v)=g_{ij}^{-1} x_j(v), 
\,\,\,\,\,\,\ R_g^{*} dx_i=g_{ij}^{-1} dx_j.
\eqn\traslada
$$
To check the horizontal character, 
notice that  $K_X$ is horizontal (for $K$ is and
$\Lambda- \theta (X_P)$ is a zero-form), and then we only have to check it for
$dx_i +\theta_{il}x_l$, as in [\mq]. Notice that we are considering $P \times
V$ as a principal bundle over $E$, and therefore a  fundamental vector field
$A^{*}$ (corresponding to $A \in {\bf g}$) is induced by the  $G$-action on
both factors. Using the
 properties of the connection one-form and the action of $G$ on $V$, one
immediately gets:
 $$
 \iota(A^{*})\theta_{ij}=A_{ij}, \,\ \iota(A^{*})dx_i= 
{\cal L}(A^{*})x_i=-A_{ij}x_j.
 \eqn\hop
 $$
 We see then that $\Phi (P\times V)$ descends to $E$. This also simplifies the
computation 
 of $d_{\hat X}\Phi (P\times V)$. First, 
 we define a connection on $P\times V$ by pulling-back 
 the connection on $P$. The horizontal subspace at $(p,v)$ is given by $H_{p}
\oplus V$, 
 where $H_p$ is the horizontal subspace of $T_pP$. If we denote by $\Phi h$
the horizontal 
 projection of a form $\Phi$ on $P \times V$ that descends to $E$, we have:
 $$
 d\Phi=d\Phi h =D\Phi,
 \,\,\,\,\,\ \iota({\hat X})\Phi =\big(\iota({\hat X})\Phi \big)h.
 \eqn\desciende
 $$
 As $\theta$ vanishes on horizontal vectors, we can put it 
 to zero after computing the exterior
 derivative of \eqtresThom. Also notice that the covariant derivative 
defined by the 
 pullback connection on $P \times V$ acts
 as the covariant derivative of $P$ on the differential
 forms in $\Omega^{*}(P)$, and as the usual exterior
 derivative on the forms in $\Omega^{*}(V)$. 
 
 Now we can compute $d_{\hat X}\Phi (P\times V)$ in a simple way. 
Again we only need to 
 compute
 the equivariant exterior derivative of the exponent:
 $$\eqalign{
 &d_{\hat X} \{ -x_i x_i
-(dx_i +\theta_{il}x_l)(K_{X})^{-1}_{ij}
(dx_j + \theta_{jm}x_m)\}\cr
&=2dx_ix_i +[K_{il}x_l-u({\cal L}(X_V)x_i+\theta_{il} x_l)]
(K_{X})^{-1}_{ij}dx_j\cr
&-dx_i[(DK_{X}^{-1})_{ij}-u\iota(X_P)(K_{X})^{-1}_{ij}]dx_j\cr
&-dx_i(K_{X})^{-1}_{ij}[K_{jp}x_p-u({\cal L}(X_V)x_j+\theta_{jp} x_p)].\cr}
\eqn\eles
$$
The computation of ${\cal L}(X_V)x_i$ is straightforward from the definition
\masflujo\ and one  obtains $-\Lambda_{ij}x_j$ as in \final. Assuming \pfbeq\
we get $DK_{X}=u\iota(X_P)K$ and therefore, using the same arguments leading
to \clarita, we see that \eles\ is zero. If we denote  by ${\tilde \pi}$ the
projection of $P \times V$ on $E$, it follows from our assumptions that 
${\tilde \pi}_{*}{\hat X} =X_E$, and therefore, using \ident\ we see that the
form induced by  \eqtresThom\ on $E$ is equivariantly closed with respect to
$X_E$.

The above computation also shows the possibility of introducing a Cartan-like
formulation of the equivariant extension we have obtained. Consider the form
on $\Omega^{*}(P\times V)[u]$  given by 
$$
 \Phi (P\times V)_{C}=(2 \pi)^{-m}{\rm Pf}(K_X){\rm exp}
\{ -x_i x_i
-dx_i(K_{X}^{\alpha})^{-1}_{ij}
dx_j \}.
\eqn\cartanlike
$$
Clearly it is still invariant under the action of $G$, but the horizontal 
character fails. However we  can consider the 
horizontal projection of this form,
$\Phi(P\times V)_{C}h$, where the horizontal subspace is defined as before by
the pullback connection. This form coincides in fact with \eqtresThom, because
the horizontal projection only applies to $dx_i$ and gives $$
(dx_i)h=dx_i+\theta_{ij} x_j.
\eqn\horizontes
$$
The interesting thing about \cartanlike\ 
is that when one enforces the horizontal
projection as in [\aj], one obtains the adequate formalism to topological
gauge theories. We will then follow this procedure to obtain a representative
which will be useful later. 

We suppose now that we have a metric $g$ on $P$ which is $G$-invariant. We use
this  metric to define the connection
 on $P$, by declaring the horizontal subspace to be the orthogonal complement
of the vertical
 one. More explicitly, one starts from the map defining fundamental vector
fields on $P$:
 $$
C_p= R_{p{*}}:{\bf g} \rightarrow T_pP.
 \eqn\funda
 $$
 Consider now the following differential form on $P$ with 
values in ${\bf g}^{*}$:
 $$
 \nu_p(Y_p, A)=g_p(R_{p{*}}A, Y_p),\,\,\,\,\ Y_p \in T_pP,\,\,\ A \in {\bf g}.
 $$
 If we denote by $C^{\dagger}_{p}$ the adjoint of $C_p$ 
(which is defined by the
 metric on $P$ together with the Killing form on ${\bf g}$), 
and let $R=C^{\dagger}C$, 
 the connection
 one-form is defined by:
 $$
 \theta=R^{-1}\nu.
 \eqn\conecta
 $$
 With the assumptions we have made concerning $P$, the condition ${\cal
L}(X_P)\theta=0$ is 
 equivalent to the metric being invariant under the vector field action. Now
we will write
 \cartanlike\ as a fermionic integral over Grassmann variables:
 $$
 \Phi (P\times V)_{C}=(\pi)^{-m}{\rm e}^{-x_ix_i}
\int {\cal D}\rho \,\ {\rm exp}
\Big({1 \over 4} \rho_i (K_X)_{ij} \rho_j +idx_i\rho_i \Big).
\eqn\fermi
$$
As we want to make a horizontal projection of this form, 
we can write $K=d\theta=R^{-1}d\nu$,
 and for the equivariant curvature defined in \claudia\ we have:
 $$
 K_X=R^{-1}(d\nu-u\nu(X_P))+u\Lambda .
 \eqn\liv
 $$
 If we introduce Lie algebra variables $\lambda$, $\phi$ 
and use the Fourier inversion formula of
 [\aj], we get the expression:
 $$\eqalign{
 \Phi (P\times V)_{C}=&(2\pi)^{-d}(\pi)^{-m}{\rm e}^{-x_ix_i}\int  {\rm exp}
\Big({1 \over 4} \rho_i (\phi_{ij}+u\Lambda_{ij}) \rho_j +idx_i\rho_i\cr
&+i\langle d\nu-u\nu(X_P), \lambda \rangle -i\langle \phi, R 
\lambda \rangle \Big){\rm det}R
\,\ {\cal D} \rho{\cal D}\phi {\cal D}\lambda, \cr}
\eqn\fourier
$$
where $\langle, \rangle$ denotes the Killing form of ${\bf g}$, 
and $d={\rm dim}\,\ G$. 
Notice that in this expression the integration 
over $\lambda$ gives a $\delta$-function constraining $\phi$ 
to be $K-\theta(X_P)$, which 
is precisely \pfbeqcurv, 
the equivariant curvature of the principal bundle $P$. To
enforce the
horizontal projection, we multiply by the normalized invariant 
volume form ${\cal D}g$ along 
the $G$-orbits, and we can write [\aj, \cmrlect]:
$$
({\rm det}R)
{\cal D}g=\int {\cal D}\eta \,\ {\rm exp}i\langle \eta, \nu \rangle ,
\eqn\medida
$$
where $\eta$ is a fermionic Lie algebra variable. 
Putting everything together we 
obtain a representative
for the horizontal projection:
$$\eqalign{
\Phi (P\times V)_{C}h=&(2\pi)^{-d}(\pi)^{-m}{\rm e}^{-x_ix_i}\int  {\rm exp}
\Big({1 \over 4} \rho_i (\phi_{ij}+u\Lambda_{ij}) \rho_j +idx_i\rho_i\cr
&+i\langle d\nu-u\nu(X_P), \lambda \rangle -i\langle \phi, R \lambda \rangle+
i \langle \eta, \nu \rangle \Big) {\cal D}\eta {\cal D}\rho {\cal D} 
\phi {\cal D} \lambda, \cr}
\eqn\lagrangiano
$$
where integration over the fibre is understood.  

We will introduce now a BRST complex in a geometrical way.
 As in the preceding section, we  introduce auxiliary fields $\pi_i$ with the
meaning of a basis of differential forms for  the fibre. The natural BRST
operator is precisely the $d_{\hat X}$ operator, but we must  take into
account that we have in $\Phi(P \times V)_{C}h$ is the horizontal projection 
of $dx_i$, given in \horizontes. Acting with the equivariant exterior
derivative and  projecting horizontally, as we did in \eles, we get: 
$$
d_{\hat X}(dx_ih)=u\Lambda_{ij}x_j +(K_{ij}-u\theta_{ij}(X_P))x_j.
\eqn\misterios
$$
Remembering that $\phi$ is equivalent to the equivariant curvature of $P$, 
the BRST operator
for the fibre is naturally given by:
$$
Q\rho_i =\pi_i, \,\,\,\,\,\,\ Q\pi_i=(u\Lambda_{ij}+\phi_{ij})\rho_j.
\eqn\alsacia
$$
Following [\cmrlect] we introduce a ``localizing" and a ``projecting" 
gauge fermion:
$$
\Psi_{\rm loc}=-\rho_i(ix_i+{1 \over 4}\pi_i),\,\,\,\,\,\,\ 
\Psi_{\rm proj}=i\langle \lambda  ,\nu \rangle.
\eqn\fermiones
$$
On the Lie algebra elements the BRST operator acts as:
$$
Q\lambda =\eta, \,\,\,\,\,\,\ Q\eta=-[\phi, \lambda].
\eqn\mascus$$
In order to obtain \lagrangiano\ from \fermiones\ using the BRST complex, we
must also  take into account the horizontal projection of forms on $P$, like
in \misterios,  and the equivariant  exterior derivative is then given as
$$
d-\iota(C\phi)-u\iota(X_P).
\eqn\masmisterio
$$
Notice that $\phi$ is an element of the Lie algebra ${\bf g}$, and therefore
$C\phi$ is  a fundamental vector field on $P$. Using \mascus\ and
\masmisterio\ as BRST operators acting on the gauge  fermions \fermiones,  the
topological lagrangian \lagrangiano\ corresponding  to an equivariant
extension of the Thom form is recovered. The BRST complex we have introduced
looks like a $G\times X_P$ equivariant  cohomology, but one shouldn't take
this analogy too seriously. If one formulates
 this equivariant cohomology  in the Weil model, the relation
$\iota(X_P)\theta =0$ should be introduced. Clearly,  this is not true 
geometrically unless $X_P$ is horizontal. In fact, this term appears in the
equivariant curvature of the principal bundle, and therefore  in the
expression for $\phi$ once the $\delta$-function  constraint has been taken
into account. 

The last point we would like to consider is the pullback of the equivariant
extension we have  obtained for this case. As \eqtresThom\ descends to a
equivariantly closed differential form on $E$, we can pull it back through an
invariant section ${\hat s}: M \rightarrow E$ as we did in subsection 3.3. But
recall that every section of $E$ is associated to a $G$-equivariant map 
$$
s:P \rightarrow V, \,\,\,\,\,\ s(pg)=g^{-1}s(p).
\eqn\eqsect
$$
If ${\hat s}$ is invariant, then  the corresponding $s$ is \eqsect\ verifies:
$$
s\phi_t^{P} = \lambda(t) s.
\eqn\dobleq
$$
Consider now the map ${\tilde s}:P \rightarrow P\times V$ given by ${\tilde
s}(p)=(p, s(p))$.  From the above it follows that  ${\tilde s}^{*}\Phi(P\times
V)$ is a closed equivariant differential form on $P$ with respect to $X_P$,
and in fact it descends to $M$, producing the same form we would get had we
used the section ${\hat s}$. We have then the conmutative diagram: 
$$
 \matrix{ \Omega^{*}(P \times V)_{{\rm basic}, E} & {\longrightarrow}  
& \Omega^{*}(E)\cr                  
                {\tilde s}^{*} \Big \downarrow & &\Big \downarrow{\hat
s}^{*}\cr
                {\Omega^{*}(P )_{{\rm basic}, M}}&{\longrightarrow} 
&\Omega^{*}(M)\cr}
\eqn\diagramaconm
$$
This diagram should be kept in mind in topological gauge theories, where the
topological  lagrangian is usually a basic form on $P$ descending to $M$. When
considering the equivariant  extension of the Mathai-Quillen form we will have
the same situation, with an equivariantly  closed differential form on $P$
descending to $M$.

\endpage

\chapter{Applications}

\section{Topological sigma models} 

Applying the previous formalism to the
topological sigma model [\topmodel] we will obtain  the model of [\llla],
which was constructed by twisting an $N=2$ supersymmetric sigma model  with
potentials [\ag]. The Mathai-Quillen formalism for usual sigma models can be
found in  [\am, \wu, \cmrlect]. 

Let $M$ be an almost hermitean manifold on which a vector field $X$ acts
preserving the almost  complex structure $J$ and the hermitean metric $G$: $$
{\cal L}(X)J={\cal L}(X)G=0.
\eqn\variedad
$$
We have then a one-parameter flow $\phi_t$ 
associated to $X$ which is almost complex with respect 
to $J$:
$$
\phi_{t*}J=J\phi_{t*}.
\eqn\casicomplejo
$$

Let $\Sigma$ be a Riemann surface with a complex structure $\epsilon$ and
metric $h$ inducing  $\epsilon$. In the topological sigma model, formulated in
the framework of the Mathai-Quillen  formalism, one takes as the base manifold
${\cal M}$ the space of maps  $$
{\cal M}={\rm Map}(\Sigma, M)=\{ f: \Sigma \rightarrow M, f \in C^{\infty}
(\Sigma, M) \}.
\eqn\base
$$
Given a $f \in {\cal M}$ we can consider the bundle over $\Sigma$ given by 
$T^{*}\Sigma \otimes 
f^{*}TM$, and define a bundle over ${\cal M}$ by 
giving the fibre on $f \in {\cal M}$:
$$
{\cal V}_f=\Gamma(T^{*}\Sigma \otimes f^{*}TM)^{+},
\eqn\fibra
$$
where $+$ denotes the self-duality constraint for the elements
 $\rho \in {\cal V}_f$:
$$
J\rho\epsilon =\rho.
\eqn\selfdual
$$ 
There is a natural way to define a vector field action on ${\cal M}$ 
induced by the action of 
$X$ on $M$:
$$
(\phi_t f)(\sigma)=\phi_t(f(\sigma)),
\eqn\arriba
$$
and similarly we can define an action on the fibre ${\cal V}_f$:
$$
({\tilde \phi}_t \rho)(\sigma)=\phi_{t*}(\rho(\sigma)).
\eqn\alafibra
$$
This action is well defined, {\it i.e.}, ${\tilde \phi}_t \rho$ verifies the
self-duality  constraint \selfdual\ when $\rho$ does, due to \casicomplejo. It
is also clear that the  compatibility conditions of
subsection 2.2 hold: first, $({\tilde \phi}_t \rho)(\sigma)$ takes values in 
$T^{*}_{\sigma}\Sigma \otimes T_{\phi_tf(\sigma)}M$, 
therefore ${\tilde \phi}_t \rho \in 
{\cal V}_{\phi_tf}$; second, the map \alafibra\ is clearly a linear map 
between fibres, as 
it is given by the action of $ \phi_{t*}$.

Now we will define metrics on ${\cal M}$ and ${\cal V}$. Let $Y$, $Z$ vector
fields on ${\cal M}$.  We can formally define a local basis on $T{\cal M}$
from a local basis on $M$, given by functional  derivatives with respect to
the coordinates: $\delta /\delta f^{\mu}(\sigma)$ [\wu].  A vector field  on
${\cal M}$ will be written locally as: 
$$
Y= \int d^2\sigma Y^{\mu}(f(\sigma)) {\delta \over \delta f^{\mu}(\sigma) }.
\eqn\campo
$$   
With respect to this local coordinate description we define the metric 
on ${\cal M}$ as:
$$
(Y,Z)=\int d^2 \sigma  {\sqrt h}G_{\mu \nu} Y^{\mu}(f(\sigma))Z^{\nu}
(f(\sigma)).
\eqn\metricaloca
$$
In a similar way, if $\rho$, $\tau \in {\cal V}_f$ have local coordinates 
$\rho^{\mu}_{\alpha}$, $\tau^{\nu}_{\beta}$, the metric on ${\cal V}_f$ 
is given by:
$$
(\rho, \tau)=\int d^2 \sigma {\sqrt h}G_{\mu \nu} h^{\alpha \beta}
\rho^{\mu}_{\alpha} \tau^{\nu}_{\beta}.
\eqn\metricafibra
$$
As $X$ is a Killing vector for the hermitean metric $G$, both \metricaloca\ 
and \metricafibra\ verify \cometrica. Now we will define a connection on
${\cal V}$ compatible  with \metricafibra. In analogy with the local basis for
$T{\cal M}$, we can construct a local basis of differential forms on
$\Omega^{*}({\cal M})$, ${\tilde d}f^{\mu}(\sigma)$, which is dual to $\delta
/\delta f^{\mu}(\sigma)$ in a functional sense: $$
({\tilde d}f^{\mu}(\sigma))\big({\delta \over \delta f^{\nu}(\sigma ')}
\big) =\delta_{\nu}^{\mu}
\delta (\sigma-\sigma ').
\eqn\dualidad
$$
Let $s$ be a section of ${\cal V}$, with local coordinates
$s_{\alpha}^{\mu}$. We will define
the 
connection by the local expression:
$$
Ds_{\alpha}^{\mu} ={\tilde d} s_{\alpha}^{\mu} +
\Big( \Gamma_{\nu \lambda}^{\mu}+{1 \over 2} 
D_{\nu}J_{\kappa}^{\mu}J^{\kappa}_{\lambda}
\Big) s_{\alpha}^{\lambda}{\tilde d}f^{\nu},
\eqn\cabraloca
$$
where ${\tilde d}$ is the exterior derivative on 
${\cal M}$, with local expression:
$$
{\tilde d} s_{\alpha}^{\mu}=\int d^2 {\sqrt h} {\delta s_{\alpha}^{\mu} 
\over \delta f^{\nu}(\sigma )}{\tilde d}f^{\nu}( \sigma).
\eqn\estereo
$$
The connection defined in this way is induced by the connection on $M$ given 
by:
$$
D=D_{G}+ {1 \over 2}D_{G}J J,
\eqn\covam
$$
where $D_G$ is the Riemannian connection canonically associated to the hermitean
metric $G$ on $M$.  Notice that, if $M$ is K\"ahler, then $D_{G}J=0$ and the
covariant derivative reduces to the usual  form. It is easy to see that
\cabraloca\ is compatible both with the self-duality constraint and with the
metric \metricafibra.

To define the usual topological sigma model we also need a section of ${\cal
V}$.  This section is essentially  the Gromov equation for pseudoholomorphic
maps $\Sigma \rightarrow M$, and can be written as:
 $$
s(f)=f_{*}+Jf_{*}\epsilon.
\eqn\seccionsigma
$$
Using \selfdual\ it is easy to show that $s$ is invariant under the vector
field action on ${\cal M}$. The last ingredient we need to construct the
equivariant extension of the  Thom form is to check the equivariance of the
connection \cabraloca. As the action of the  vector field $X$ on ${\cal M}$ is
induced by the corresponding action on $M$, it is  sufficient to prove the
equivariance of the connection \covam\ (equivalently, if we check the 
equivariance in local coordinates for ${\cal M}$, ${\cal V}$, we are reduced
to a computation  involving the local coordinate expressions of $X$ and $D$ on
$M$). If $X$ is a Killing vector  field for the metric $G$ one has ${\cal
L}(X)D_G=D_{G}{\cal L}(X)$. Using now \variedad\ it is  clear that ${\cal
L}(X)$ conmutes with $D$, hence $D$ is equivariant and also the connection on 
${\cal V}$ defined in \cabraloca.

Therefore , we are in the conditions of subsection 3.3, and   we can construct
the equivariant extension of the Thom form introduced there. To do this  we
must first of all compute the operator $L_{\Lambda}=\Lambda-\theta(X)$ in
local coordinates.  As before, the computation reduces to a local coordinate
computation on the  target manifold $M$. Fist we will obtain $\Lambda$ through
the equation \final. Take as local  coordinates on the fibre
$\rho_{\alpha}^{\mu}(\sigma) $. We have: 
$$
({\cal L}(X)\rho_{\alpha}^{\mu})(\sigma)=\lim_{t \rightarrow 0}
{(\phi_{t*}\rho)_{\alpha}^{\mu}(\sigma)-\rho_{\alpha}^{\mu}(\sigma) \over t} =
\lim_{t \rightarrow 0}{1 \over t}\Big( 
{\partial (u^{\mu}\phi_t) \over \partial u^{\nu}}-
\delta_{\nu}^{\mu} \Big) \rho_{\alpha}^{\nu}(\sigma),
\eqn\maslie
$$
where  $u^{\mu}$ are local coordinates on $M$ and we explicitly wrote the
jacobian matrix  associated to $\phi_{t*}$. The limit above is easily computed
once we take into account  that the one-parameter flow in local coordinates 
$(u^{\mu}\phi)(t, u)= g^{\mu}(t,u)$ verifies the differential system: 
$$
{\partial g^{\mu}(t,u) \over \partial t} =X^{\mu}(g(t,u)), 
\,\,\,\,\,\ g^{\mu}(0,u)=u^{\mu},
\eqn\sistema
$$
where $X^{\mu}(g(t,u))$ is the local coordinate 
expression of the vector field $X$ 
associated to the flow. Using \sistema\ we get:
$$
({\cal L}(X)\rho_{\alpha}^{\mu})(\sigma)=
(\partial_{\nu} X^{\mu})(f(\sigma)) \rho_{\alpha}^{\nu}(\sigma).
\eqn\otrolio
$$
Taking into account that the indices for local 
coordinates on ${\cal V}_f$ are $\mu$, $\alpha$, 
we finally obtain:
$$
\Lambda^{\mu \alpha}_{\nu \beta}(f(\sigma))=-(\partial_{\nu} 
X^{\mu})(f(\sigma)) 
\delta_{\beta}^{\alpha}.
\eqn\mimatriz
$$
Next we compute $\theta(X)$. Again, by \cabraloca, we can compute 
it for the connection 
matrix on $M$ given by \covam:
$$
\theta =\theta_G+ {1 \over 2}D_{G}J J,
\eqn\conexion
$$
where $\theta_G$ is the Levi-Civita connection associated to the metric $G$. 
Using \variedad\ we get:
$$
\theta (X) ={1 \over 2} \Big( \theta_G (X)-J\theta_G(X)J \Big), 
\eqn\contraete
$$
To obtain the additional term in the topological action \campos\ corresponding
to the  operator $L_{\Lambda}$, we must act on coordinate fields for the fibre
which are self-dual  and verify \selfdual. Using this constraint it is easy to
see that \contraete\ is  equivalent to $\theta(X)$. We can already write this
term $u\rho_i (L_{\Lambda})_{ij} \rho_j/4$  as
$$
\int_{\Sigma} d^2 \sigma {\sqrt h} {u \over 4} 
h^{\alpha \beta} \rho^{\nu}_{\beta}
D_{\nu}X_{\mu}\rho^{\mu}_{\alpha},
\eqn\adicion
$$
where $D_{\nu}$ is the Levi-Civita covariant derivative on $M$, and we have
used the  Grassmannian character of the fields $\rho$. This is precisely the
extra term obtained  in [\llla] after the twisting of the $N=2$ supersymmetric
sigma model with potentials. 

In the  topological action of [\llla] there are
also two additional terms that in the topological  model come from a $Q$-exact
fermion and have a counterpart in the non-twisted action. Remarkably, these
two terms can be interpreted as the $d_{X}$-exact equivariant differential
form  that is added to prove localization in equivariant integration [\berlin,
\blau]. We will present the general setting and  then apply it to the
equivariant extension of the topological sigma model. As we will see,  the
same construction holds for non-abelian monopoles on four-manifolds. Notice 
that, this additional term being $d_{X}$-exact, we can multiply it by an 
arbitrary parameter $t$ without changing the equivariant cohomology class. 
This can be exploited to give saddle-point-like proof of 
localization of equivariant integrals on the critical
 points of the vector field action (or, equivalently, on 
the fixed points of the associated one-parameter action). 
Suppose then that on the base manifold $M$ there is a metric $G$ and that the
vector field $X$ acts as a Killing vector field with respect to $G$. Consider
the differential form given by 
$$
\omega_{X}(Y)=G(X,Y),
\eqn\laforma
$$ $Y$ a vector field on $M$. As $X$ is Killing, 
we have ${\cal L}(X) \omega_X=
0$, and acting with $d_{X}$ gives the equivariantly exact differential form 
$$
d_{X}\omega_{X}=d\omega_{X}-uG(X,X).
\eqn\exacta
$$
The appearance of the norm of the vector field $X$ in \exacta\ is what gives 
localization on the critical points of the vector field. In the topological
 sigma model there is a metric on ${\cal M}$ 
given in \metricaloca\ which 
is Killing with respect to the action of $X$ on ${\cal M}$, and therefore we 
can add the exact form \exacta\ to our 
equivariantly extended topological action. In fact \laforma\ is explicitly 
given on ${\cal M}$ by:
$$
\omega_{X}=\int d^2 \sigma  {\sqrt h}G_{\mu \nu} X^{\mu}(f(\sigma))
{\tilde d}f^{\nu}(\sigma).
\eqn\explicito
$$
We can then obtain \exacta\ in this case as
$$
d_{X} \omega_{X} =\int d^2 \sigma {\sqrt h}\Big(\chi^{\mu} 
\chi^{\nu} D_{\mu}X_{\nu}-
uG_{\mu \nu} X^{\mu}X^{\nu} \Big),
\eqn\masterminos
$$
where we have introduced the usual field theory representation of the basis of 
differential forms, $\chi^{\mu}= {\tilde d}f^{\mu}$. With \adicion\ and
\masterminos\ we recover all the terms of the sigma model of [\llla] beside 
the usual ones. The BRST complex for the equivariant extension of the
topological sigma model  follows from our indications in subsection 3.3, and
coincides with the one in [\llla] after a  redefinition of the auxiliary
fields, as we will see in sect. 5. 
As a last remark, notice that the observables of this  topological
field theory are naturally associated to the equivariant cohomology classes
on  $M$ with respect to the action of $X$. The equivariant extension of the
topological sigma  model is thus the natural framework to study quantum
equivariant cohomology. 

\section{Non-abelian monopoles on four-manifolds}

Non-abelian monopole equations on four-manifolds were introduced in [\lm], in
the  framework of the Mathai-Quillen formalism, as a generalization of
Donaldson-Witten theory  [\don, \donfirst, \donkron, \tqft] and  of the
Seiberg-Witten abelian monopole equations [\ws, \mfm]. Other studies of these
equations  can be found in [\ansdos-\oscar, \korea].
 From the physical point of view,
these models can be  understood as twisted $N=2$ Yang-Mills theories coupled
to massless  matter hypermultiplets [\alab,  \alabas, \ans, \tqcd], and this
fact in turn allows a computation of the associated  topological invariants
using non-perturbative results for supersymmetric gauge theories  [\wjmp,
\mfm, \lmdos]. We will exploit  the fact that the model has a $U(1)$ symmetry
[\pids, \tqcd, \korea] to obtain an equivariant  extension of the Thom form in
this case. We will obtain a theory which corresponds  to a twisted $N=2$
Yang-Mills theory coupled to massive matter multiplets.  The connection
between the $U(1)$ equivariant cohomology and the massive theory was  pointed
out in [\korea]. 

Non-abelian monopoles on four-manifolds are described by a topological gauge
theory, and then  we will follow the general procedure in subsection 3.4
above. The geometrical data of the  theory are as follows [\lm].  
 Let $X$ be an oriented, compact four-manifold endowed with a Riemannian
structure given by a metric $g$. We will restrict ourselves to spin manifolds,
although the generalization to arbitrary manifolds can be done using a ${\rm
Spin}_c$ structure . We will denote the positive and negative chirality spin
bundles on $X$ by $S^{+}$ and $S^{-}$, respectively. We also consider on $X$ a
principal fibre bundle $P$ with some compact, connected, simple Lie group $G$.
The Lie algebra of $G$ will be denoted by ${\bf g}$. For the matter part we
need an associated vector bundle $E$ to the principal bundle $P$ by  means of
a representation $R$ of the Lie group $G$. Now, for the principal bundle of
the moduli problem (not to be confounded with $P$), we consider ${\cal
P}={\cal A} \times \Gamma (X, S^{+} \otimes E)$, where ${\cal A}$ is the
moduli space of $G$-connections on $E$, and $\Gamma (X, S^{+} \otimes E)$ is
the space of sections of the bundle $S^{+}\otimes E$. As the group ${\cal G}$
acting on this principal bundle we take the group of gauge transformations of
the bundle $E$, whose action on the moduli space is given locally by:    
$$ 
\eqalign{ &g^{*}(A_\mu)=-igd_\mu g^{-1}+gA_\mu g^{-1},\cr
&g^{*}(M_{\alpha})=g M_{\alpha},\cr }   \eqn\gaugedos  
$$   
where $M \in \Gamma (X, S^{+} \otimes E)$ and $g$ takes values in the group
$G$ in the representation $R$. Notice that, as usual in gauge theories, we
suppose  that the gauge group acts on ${\cal P}$ on the left. As the fibre we
take the  (infinite-dimensional) vector space ${\cal F}=\Omega^{2,+} (X, {\bf
g}_E) \oplus  \Gamma (X, S^{-} \otimes E)$, where $\Omega^{2,+} (X, {\bf
g}_E)$ denotes the self-dual differential forms on $X$ taking values in the
representation of the Lie algebra of $G$ associated to $R$, ${\bf g}_E$. The
group of gauge transformations acts on ${\cal F}$ in the obvious way.
 The Lie algebra of the group ${\cal G}$ is Lie$({\cal G})=\Omega^0(X,{\bf
g}_E)$. The tangent space to the moduli space at the point $(A,M)$ is just
$T_{(A,M)}{\cal M}=T_{A}{\cal A} \oplus T_{M}\Gamma (X, S^{+} \otimes
E)=\Omega^{1}(X,{\bf g}_E) \oplus \Gamma (X, S^{+} \otimes E)$, for $\Gamma
(X, S^{+} \otimes E)$ is a vector space.   We can define a gauge-invariant
Riemannian  metric on ${\cal P}$ given by:    
$$  
g_{\cal P}\big( (\psi, \mu), (\theta, \nu) \big)=\int_{X} \tr(\psi
\wedge *\theta) +{1 \over 2} \int_{X} e ({\bar \mu}^{\alpha i}
\nu_{\alpha}^i+ \mu_{\alpha}^i {\bar \nu}^{\alpha i}),  
\eqn\pina 
$$   
where $e=\sqrt g$. The spinor notation follows that in [\lm]. 
An analogous expression gives the inner product on the fibre ${\cal F}$. The
Lie algebra of the gauge group of transformations ${\rm Lie}({\cal G})$ is
also endowed with a metric given, as in \pina, by the trace and the inner
product on the space of zero-forms. For simplicity we will  take $G=SU(N)$ and
the monopole fields $M_{\alpha}$ in the fundamental representation of this 
group.

Now we define vector field actions on ${\cal P}$ and ${\cal F}$
associated to a $U(1)$ action 
as follows:
$$
\eqalign{
\phi_t^{\cal P} (A, M_{\alpha} ) =& (A, {\rm e}^{it}M_{\alpha}), \cr
\phi_t^{\cal F} (\chi, M_{\dot \alpha}) =& (\chi, {\rm e}^{it}M_{\dot\alpha}), 
\cr}
\eqn\circulos
$$
where $M_{\alpha} \in \Gamma
(X, S^{+} \otimes E)$, $M_{\dot \alpha} \in \Gamma
(X, S^{-} \otimes E)$ and $\chi \in \Omega^{2,+} (X, {\bf g}_E)$. 
It is clear that these actions conmute with the action of the group of gauge 
transformations on both ${\cal P}$ and ${\cal F}$. Furthermore, the metrics
 on these
spaces are preserved by the $U(1)$ action. 
The section $s:{\cal P} \rightarrow {\cal F}$ defining the 
non-abelian monopole equations is:
$$
s(A,M)=\Big({1 \over \sqrt 2} \big(F^{+ ij}_{\alpha \beta}+{i\over 2}
({\overline M}_{(\alpha}^j  M_{\beta )}^i-{\delta^{ij}\over N} {\overline
M}_{(\alpha}^k  M_{\beta )}^k)\big), (D_{\alpha {\dot \alpha}
}M^{\alpha})^i\Big),   
\eqn\seccion
$$
and is clearly equivariant with respect 
to the $U(1)$ actions given in \circulos. Namely,
$$
s\big(\phi_t^{\cal P} (A, M_{\alpha} )\big)=\phi_t^{\cal F}s(A, M_{\alpha}).
\eqn\equimono$$
We are in the conditions of subsection 
3.3, and therefore we can construct the equivariant 
extension of the Thom form of 
the associated 
vector bundle ${\cal E}={\cal P} \times {\cal F}/ {\cal G}$. 
First we compute the 
${\Lambda}$ matrix on the fibre according to \openuevo. 
In local coordinates we get:
$$
\Lambda \chi =0, \,\,\,\,\,\,\ \Lambda M_{\dot \alpha}^{j}=
-iM_{\dot \alpha}^{j}.
\eqn\lieataca
$$ Notice that, if we split $M_{\dot \alpha}^{j}$ in its real 
and imaginary parts, 
$\Lambda$ is given by the matrix \matriz. From \circulos\ and 
\vector\
we can also obtain the local 
expression of the associated vector field $X_{{\cal P}}$ in $(A,M_{\alpha})$:
$$
X_{\cal P}=(0, iM_{\alpha}) \in \Omega^{1}(X,{\bf g}_E) \oplus 
\Gamma (X, S^{+} \otimes
E).
\eqn\campillo$$ 
The additional terms we get in the topological lagrangian \lagrangiano\ after
the equivariant  extension are associated to $\Lambda$, which has already been
computed, and to $\nu(X_P)$.  The explicit expression of $\nu$ was obtained in
[\lm]. For $G=SU(N)$ and the monopole  fields in the fundamental
representation it reads: 
$$
\nu(\psi,\mu)^{ij}=-(d_A^{*}\psi)^{ij} + {i \over 2}\big({\bar
\mu}^{\alpha j}M_{\alpha}^i-{\overline M}^{\alpha
j}\mu_{\alpha}^i-{\delta^{ij}\over N} ({\bar \mu}^{\alpha
k}M_{\alpha}^k-{\overline M}^{\alpha k}\mu_{\alpha}^k)\big) \in
\Omega^{0}(X,{\bf g}_E), \eqn\marga  
$$   
where $(\psi,\mu) \in
T_{(A,M)}{\cal P}$. 
Using now \marga\ and \campillo\ we get:
$$
\nu(0, iM^{i}_{\alpha})={\overline
M}^{\alpha j}M_{\alpha}^i-{\delta^{ij}\over N} {\overline M}^{\alpha
k}M_{\alpha}^k.
\eqn\masexplicito
$$
The additional terms in the topological lagrangian due to the equivariant 
extension are 
then given by:
$$
u \int_{X} e \big( -{i \over 4}{\bar v}^{\dot\alpha} v_{\dot\alpha}-
i{\overline M}^{\alpha}\lambda M_{\alpha} \big)
\eqn\masillas
$$
where we have deleted the $SU(N)$ indices, and $v_{\dot\alpha}$ is the
auxiliary field  associated to the monopole coordinate on the fibre [\lm]. The
BRST cohomology of the resulting model  was also indicated in  subsection 3.4.
Not all the terms coming from the twisting  of the massive multiplet appear,
but we can add a $d_{X_{\cal P}}$-exact piece to the action starting  with a
differential form like the one in \laforma. Now we must take into account that 
we can only add to the topological lagrangian basic forms on ${\cal P}$ which
descend to  ${\cal P}/{\cal G}$. If we define a differential form on ${\cal
P}$ starting from \pina\  as
$$
\omega_{X_{\cal P}}(Y)=g_{\cal P}(X_{\cal P}, Y),
\eqn\formilla
$$
we can use invariance of the $g_{\cal P}$ and $X_{\cal P}$ under the action of
the gauge group  to see that the above form is in fact invariant. But the 
horizontal character of \formilla\ is only guaranteed  
if $X_{\cal P}$ is horizontal.
This is in fact not true in our case, as it follows from  \masexplicito.
Therefore we must enforce a horizontal projection of $\omega_{X_{\cal P}}$
using  the connection on ${\cal P}$, and consider the form
$\omega^{h}_{X_{\cal P}}= \omega_{X_{\cal P}}h$. Actually we are interested
in  
$$ d_{X_{\cal P}}\omega^{h}_{X_{\cal P}}=d\omega^{h}_{X_{\cal P}}-
u\iota(X_{\cal P})\omega^{h}_{X_{\cal P}},
\eqn\derivando
$$
which also descends to ${\cal P}/{\cal G}$. In computing the above equivariant
exterior  derivative we must be careful, as in \misterios. This can be easily
done using the BRST  complex that we motivated geometrically in \alsacia\ and
\masmisterio. Of course, from \pina\  and \campillo\ we can give an explicit
expression of \formilla. Introducing a  basis of differential  forms for
$\Gamma(X, E \otimes S^{+})$, we get: 
$$
\omega_{X_{\cal P}}={i \over 2} \int_{X} e\big({\bar \mu}^{\alpha
}M_{\alpha}-{\overline M}^{\alpha }\mu_{\alpha}\big).
\eqn\masmasilla
$$
Acting with $d_{X_{\cal P}}$ or, equivalently, with the BRST operator, we get:
$$
Q\omega_{X_{\cal P}}=-i \int_{X} e{\bar \mu}^{\alpha
}\mu_{\alpha}-\int_{X} e{\overline M}^{\alpha }\phi M_{\alpha}- 
u\int_{X} e {\overline M}^{\alpha } M_{\alpha}.
\eqn\masotta$$
As we will see, with \masillas\ and \masotta\ we 
reconstruct all the terms appearing in the 
twisted theory with a massive hypermultiplet.

The observables in Donaldson-Witten theory and in the non-abelian monopole 
theory are differential forms on the corresponding moduli spaces, and they 
are constructed from the  horizontal projections of differential forms on the
principal bundle associated to the  problem. They involve the curvature form
of this bundle. In the equivariant extension  of the monopole theory these
observables have the  same form, but one must use instead the equivariant
curvature of the bundle, given in \pfbeqcurv.  From the point of view of the
BRST complex they have the usual form of Donaldson-Witten  theory: 
$$
{\cal O}={1 \over 8\pi^2}\tr 
\phi^2,\,\,\,\,\ I(\Sigma)={1 \over 8\pi^2}\int_{\Sigma}
\big(\phi F+{1 \over 2} \psi \wedge \psi \big),
\eqn\observa
$$
where $F$ is the Yang-Mills field strength, $\psi$ represent a basis of
differential forms  on ${\cal A}$, and $\phi$ is the Lie algebra variable
introduced in \fourier. As we have pointed out,  the $\delta$-function
involved in \lagrangiano\ constrains $\phi$ to be the equivariant curvature 
of the bundle ${\cal P}$, $K_{X_{\cal P}}$. To check that the forms in
\observa\ are closed one must be careful with  the horizontal projection
involved in the computation. Although the vector field $X_{\cal P}$ doesn't
act  on ${\cal A}$, the contraction $\iota (X_{\cal P})\psi$ is not zero, as
$\psi$ must be horizontally  projected and the field $X_{\cal P}$ must be
substituted by  $X_{\cal P}h=X_{\cal P}-R_{p*}\theta (X_{\cal P})$. 
Of course, using
the BRST complex this verification is automatic, but one should not forget the
geometry hidden inside it.

\endpage

\chapter{Twisting $N=2$ supersymmetric theories with a central charge} 

The aim of this section is to show that the topological quantum field theories
obtained in the previous section can be obtained after twisting $N=2$
supersymmetric theories having as a common feature the presence of a non-trivial
central charge. There are several reasons to believe that topological quantum
field theories resulting from the equivariant extension of the Mathai-Quillen
formalism are intimately related to twisted $N=2$ supersymmetric theories with
a non-trivial central charge. First, as we will discuss below, twisted
$N=2$ supersymmetric theories with a non-trivial central charge have the same
right to lead to topological quantum field theories as the ones with a trivial
central charge. Second, the presence of a non-trivial central charge can be
regarded as the existence of a global $U(1)$ symmetry with a structure
very much alike the gauge structure appearing in twisted $N=2$ supersymmetric
Yang-Mills theory or Donaldson-Witten theory, in clear analogy with the
structure uncovered in the previous sections.

In this section we will first develop
these general features and then we will describe in two subsections how they
are realized in two and four dimensions after considering topological sigma
models with potentials and $N=2$ supersymmetric Yang-Mills theory coupled to
massive $N=2$ supersymmetric matter fields. We will conclude that indeed the
resulting topological quantum field theories are the ones constructed in the
equivariant extension of the Mathai-Quillen formalism of the previous section.
As already indicated in that section, the  resulting two-dimensional field
theory was first constructed in [\llla] from the perspective of building a
generalization of topological sigma models. The four-dimensional topological
quantum field theory was first presented in [\korea]. In the present work
we will emphasize the role played by the non-trivial central charge in the
construction of this theory 
from the point of view of twisting $N=2$ supersymmetry.

Let us begin reviewing the standard arguments which indicate that topological
quantum field theories can be obtained after twisting $N=2$ supersymmetric
theories. We will concentrate first in $d=4$. In ${\bf R^4}$ the global symmetry group of $N=2$ supersymmetry is
${\cal H}=SU(2)_L\otimes SU(2)_R\otimes SU(2)_I\otimes U(1)_{\cal R}$ where
${\cal K}=SU(2)_L\otimes SU(2)_R$ is the rotation group, and $SU(2)_I$ and
$U(1)_{\cal R}$ are internal symmetry groups. The supercharges $Q_\alpha^i$
and ${\bar Q}_{\dot\alpha i}$ of $N=2$ supersymmetry transform under ${\cal
H}$ as $(1/2,0,1/2)^1$ and $(0,1/2,1/2)^{-1}$, respectively, and satisfy:
$$
\eqalign{
\{ Q_\alpha^i,{\bar Q}_{\dot\beta j} \} =  & \delta^i_j
P_{\alpha\dot\beta}, \cr
\{ Q_\alpha^i,Q_{\beta}^j \} =  & \epsilon^{ij} C_{\alpha\beta} Z, \cr}
\eqn\apple
$$
where $\epsilon^{ij}$ and $C_{\alpha\beta}$ are $SU(2)$ invariant tensors,
and $Z$ is the central charge generator. The twist consists of considering
as the rotation group the group ${\cal K}'=SU(2)_L'\otimes SU(2)_R$ where 
$SU(2)_L'$ is the diagonal subgroup of $SU(2)_L\otimes SU(2)_I$. Under the new
global symmetry group ${\cal H}'= {\cal K}'\otimes U(1)_{\cal R}$ the
supercharges transform as $(1/2,1/2)^{-1}\oplus (1,0)^1 \oplus (0,0)^1$. The
twisting is achieved replacing any isospin index $i$ by a spinor index
$\alpha$ so that $Q_\alpha^i \rightarrow Q_{\alpha}{}^\beta$ and
$\bar Q_{\dot\beta i} \rightarrow G_{\alpha \dot\beta}$. The $(0,0)^1 $
rotation  invariant operator is $Q=Q_{\alpha}{}^\alpha$ 
and satisfies the twisted
version of the $N=2$ supersymmetric algebra \apple, often called topological
algebra:
$$
\eqalign{
\{ Q,G_{\alpha\dot\beta} \} =  & P_{\alpha\dot\beta}, \cr
\{ Q,Q\} =  & Z. \cr}
\eqn\orange
$$
In a theory with trivial central charge the right hand side of the last of
these relations effectively vanishes and one has the ordinary situation in
which $Q^2=0$. The first of these relations is at the heart of the standard
argument to conclude that the resulting twisted theory will be topological.
Being the momentum tensor $Q$-exact it is likely that the whole
energy-momentum tensor is $Q$-exact. This would imply that the vacuum
expectation values of $Q$-invariant  operators which do not involve the
metric  are metric independent, \ie, that the theory is topological. To our
knowledge, all the twisted $N=2$ theories which have been studied
satisfy this property. The important point to remark here is that in the
presence of a non-trivial central charge the first relation in \orange\ holds
and therefore one has the same expectations to obtain a topological quantum
field theory as in the ordinary case.

The central charge generator enters in the second relation in $\orange$.
We are familiar with the presence of similar relations in Donaldson-Witten
theory. Indeed, as it is well known, the supersymmetric theories involving
Yang-Mills fields close the supersymmetric algebra up to a gauge
transformation. This implies that in a twisted theory one does not have
that $Q^2$ vanishes but that it is a gauge transformation. This is the case of
Donaldson-Witten theory in which the gauge parameter on the right hand side of
the equation for $Q^2$ is one of the scalar fields of the theory, and one is
then instructed to consider gauge invariant operators which are $Q$-invariant as
the observables of the theory. In that situation, since gauge invariant 
operators
which are $Q$-exact lead to vanishing vacuum expectation values one has to deal
with the corresponding equivariant cohomology. In this framework one can regard
the second relation in \orange\ as a situation similar to the case of
Donaldson-Witten theory where the gauge symmetry is a global $U(1)$ symmetry. In
addition, this analogy implies that the  correct mathematical framework to
formulate these theories must involve an equivariant extension.

The realization of topological quantum field theories coming from twisted
$N=2$ supersymmetric theories with a non-trivial central charge is very
interesting. Recall that in the four-dimensional case non-trivial central
charges appear when there are massive particles. This means that the resulting
topological quantum field theory is likely to possess a non-trivial parameter.
In other words, it is likely that the vacuum expectation values of its
observables, \ie, the topological invariants, are functions of this parameter.
This is a very surprising feature, specially if one thinks that the origin of
that parameter is a mass, but, at the same time, very appealing. Recall that in
ordinary Donaldson-Witten theory as well as in its extensions involving
twisted massless matter fields the action of the theory turns out to be
$Q$-exact and therefore no dependence on the gauge coupling constant
appears in the vacuum expectation values. As it will be clear below, in the
presence of a non-trivial central charge the action can again be written in a
$Q$-exact form and therefore there is no dependence on the gauge coupling
constant. However, one can not argue so simply independence of the 
parameter  originated from the mass or central charge of 
the physical theory. In this case the parameter not only enters in the $Q$-exact action but also in
the $Q$-transformations. Notice that vacuum expectation 
values in these topological theories should be 
interpreted as integrals of 
equivariant extensions of differential forms. From the 
equivariant 
cohomology point of view, the parameter 
of the central charge is the generator of the 
cohomology ring, which we have 
denoted by $u$, and the integration of
an equivariant extension of a differential form can give additional 
contributions because of the new terms needed in the extension. These 
contributions have the form of a polynomial in $u$. Therefore, we should expect 
a dependence of the vacuum expectation values of the twisted theory with 
respect to this parameter. A different 
situation arises when one considers the addition of  equivariantly exact 
forms like \exacta\ or \derivando\ multiplied by another parameter $t$. If some 
requirements of compactness 
are fulfilled, the topological invariants don't depend on this $Q$-exact piece, 
and we can compute them for different values of $t$. 
This is precisely the usual 
way to prove localization of equivariant integrals. It is likely that a 
rigorous application of this method to the models considered in this paper 
can provide new ways to compute the corresponding topological invariants. 

In ${\bf R}^2$ the global symmetry group of $N=2$ supersymmetry is ${\cal
H}=SO(2)\otimes U(1)_L\otimes U(1)_R$ where ${\cal
K}=SO(2)$ is the rotation group, and $U(1)_L$ and $U(1)_R$ are left and right
moving chiral symmetries. There are four supercharges $Q_{\alpha a}$ 
transforming
under ${\cal H}$ as $(-1/2,1,0)$, $(-1/2,-1,0)$,
$(1/2,0,1)$ and $(1/2,0,-1)$. They  satisfy: 
$$ 
\eqalign{ \{ Q_{\alpha +},Q_{\beta -} \} =  & 
P_{\alpha\beta}, \cr 
\{ Q_{\alpha +},Q_{\beta +} \} = &
\{ Q_{\alpha -},Q_{\beta -} \} = \epsilon_{\alpha\beta} Z, \cr} 
\eqn\appledos
$$
where  $\epsilon_{\alpha\beta}$ is an antisymmetric $SO(2)$ invariant tensor,
and $Z$ is the central charge generator. The twist consists of considering
as the rotation group the diagonal subgroup of $SO(2)\otimes SO(2)'$,
where $SO(2)'$ has as generator $(U_L-U_R)/2$ being $U_L$ and $U_R$ the
generators of $U(1)_L$ and $U(1)_R$ respectively. Under the new global symmetry
group ${\cal H}'= SO(2) \otimes U(1)_{F}$, where $U(1)_{F}$ has as generator
the combination $U_L+U_R$, the supercharges transform as 
$(0,1)\oplus (-1,-1) \oplus (0,1) \oplus (1,-1) $. The twisting is achieved
thinking of the second index of $Q_{\alpha a}$ as an $SO(2)$ isospin index and,
as in the four dimensional case, replacing any isospin index $a$ by a spinor
index $\beta$ so that $Q_{\alpha a} \rightarrow Q_{\alpha\beta}$. 
One of the two  rotation  invariant operators is $Q=Q_{\alpha}{}^\alpha$. It
satisfies the twisted version of the $N=2$ supersymmetric algebra \appledos\ or 
topological algebra:  
$$ \eqalign{
\{ Q,G_{\alpha\beta} \} =  & P_{\alpha\beta}, \cr
\{ Q,Q\} =  & Z, \cr}
\eqn\orangedos
$$
where $G_{\alpha\beta}$ is the symmetric part of $Q_{\alpha\beta}$.
Notice that one could have taken  the 
combination $(U_L+U_R)/2$ instead of $(U_L-U_R)/2$ in order to carry out the
twisting. This would have led to the second type of twisting  discussed in 
 [\egu,\tmintwo,\mima]. However, as shown in [\tmintwo],
the twisting of $N=2$ supersymmetric chiral multiplets and twisted chiral
multiplets is interchanged by the two types of twisting. Thus without loss of
generality we can restrict ourselves to one type of twist since, as it becomes
clear below, we will discuss the aspects of the twist of the two types of $N=2$ 
supersymmetric multiplets.

In the two-dimensional case the central charge generator of $N=2$
supersymmetry acts as a Lie derivative with respect to a Killing vector field. 
This feature holds in the twisted theory for the right hand side of the
expression for $Q^2$. This implies, on the one hand, that the theory exists for
a restricted set of target manifolds as compared to the ordinary
topological sigma models. On the other hand, the theory is very interesting
because, as in the case in four dimensions,  one finds  topological
invariants which are sensitive to the kind of Killing vector chosen, and one
might discover new ways to compute topological invariants. 


\section{Topological sigma models with potentials}

We  begin recalling a few standard facts on non-linear sigma models in two
dimensions. Non-linear sigma models involve mappings from a two-dimensional
Riemann surface $\Sigma$ to an $n$-dimensional target manifold $M$.
The local coordinates of this mapping can be regarded as bosonic
two-dimensional fields which might be part of different types of supersymmetric
multiplets. In $N=2$ supersymmetry there are two types of multiplets, chiral
multiplets and twisted chiral multiplets. The possible geometries of the target
manifold $M$ are severely restricted by the different choices of multiplets
taking part of a given model. In models involving only chiral multiplets $N=2$
supersymmetry requires that $M$ is a K\"ahler manifold 
[\zumino,\luisf]. In the situations where both multiplets are allowed, $M$ can
be a hermitean locally product space 
[\martin]. Twistings of models involving both types of multiplets have been
considered in [\topmodel,\vafa,\tmintwo,\mima]. 

We will concentrate in the case in which there are only
chiral multiplets. 
Twisted chiral multiplets lead to topological quantum field theories which are
not well suited to be reformulated in the Mathai-Quillen formalism.
The case of chiral multiplets was the one considered by E. Witten when
topological sigma models were formulated for the first time [\topmodel]. As
shown in [\topmodel] it turns out that after the twisting the constraint
present in the $N=2$ supersymmetric theory which imposes $M$ to be K\"ahler can
be relaxed and it turns out that the twisted model exists for target manifolds
which are almost hermitean. This fact is not surprising since in the
topological theory one demands only the existence of one half of a
supersymmetry out of the two supersymmetries which are present before the
twisting. However, the twisting of the most general $N=2$ supersymmetric
theory involving only chiral multiplets was not considered in [\topmodel]. As
shown in [\ag] potential terms can be introduced for $N=2$ supersymmetric sigma
models. It was shown in [\tmintwo] that the  potential terms which appear
through $F$-terms are not allowed
because they are inconsistent with Lorentz invariance after the twisting.
However, the other types of potential terms contained in the formulation
presented in [\ag] are permitted. These potential terms only exist for
manifolds which admit at least one holomorphic Killing vector field.
The twisting of these models leads to the topological quantum field
theory constructed in the previous section.

Twisted $N=2$ supersymmetric sigma models with potential terms associated to
holomorphic Killing vectors have been considered in [\llla]. As in the
ordinary case, the K\"ahler condition on $M$ can be relaxed and the topological
model exist for any almost hermitean manifold admitting at least one
holomorphic Killing vector. Although most of what comes out in our analysis
is already in [\llla], we will describe the construction in some detail to
point out the close parallelism with the situation in four dimensions.

Let $M$ be a $2d$-dimensional K\"ahler manifold endowed with a hermitian metric
$G$ and a complex structure $J$. This complex structure verifies
$D_\rho J^\mu{}_\nu=0$, where $D_\rho$ is the covariant derivative
with the Riemann connection canonically associated to the hermitian metric
$G$ on $M$.
The action which results after performing the twist of
the $N=2$ supersymmetric action given in [\ag] (with the functions $h$ and
$G^i$ set to zero) is [\llla]:
$$
\eqalign{
S_1 =& \int_\Sigma d^2z \sqrt{h}
\Big[{1\over 2} G_{\mu\nu}h^{\alpha\beta}\partial_\alpha x^\mu \partial_\beta
x^\nu - ih^{\alpha\beta}G_{\mu\nu}\rho^\mu_\alpha D_\beta\chi^\nu 
-{1\over 8}h^{\alpha\beta}
R_{\mu\nu\sigma\tau}\rho^\mu_\alpha\rho^\nu_\beta\chi^\sigma\chi^\tau \cr &
\,\,\,\,\,\,\,\,\,\,\,\,\,\,\,\,\,\,\,\,\,\,\,\, + G_{\mu\nu}X^\mu X^\nu -
\chi^\mu\chi^\nu D_\mu X_\nu - {1\over 4}h^{\alpha\beta}
\rho_\alpha^\mu\rho^\nu_\beta D_\mu X_\nu\Big], \cr}
\eqn\dos
$$
where $h$ is the metric on the Riemann surface $\Sigma$. In the
action \dos, $x^i$, $i=1,\dots,2d$, are bosonic fields which 
describe locally
a map $f: \Sigma \rightarrow M$, and  $\rho_\alpha^\mu$,
$i=1,\dots,2d$, are anticommuting fields which are sections of ${\cal V}_f$
in \fibra. The fields $\rho_\alpha^\mu$
satisfy the self-duality condition, $\rho_\alpha^\mu = \epsilon_\alpha{}^\beta
J^\mu{}_\nu \rho^\nu_\beta$, in \selfdual. 
The fields $\chi^\mu$ constitute a basis of differential forms 
$\tilde d f^\mu$.
In  \dos\ $D_\alpha$ is the pullback
covariant derivative (eq. \cabraloca\ in the K\"ahler case, $D_\rho
J^\mu{}_\nu=0$) and $X^\mu$ is a holomorphic Killing vector field on $M$
which besides preserving the hermitean metric $G$ on $M$ it also preserves the
complex structure $J$. These two features are contained in the conditions
\variedad\ which are equivalent to:
$$
D_\mu X_\nu + D_\nu X_\mu =0,
\,\,\,\,\,\,\,\,\,\,\,\,
J^\mu{}_\nu J^\nu{}_\rho  D_\mu X_\nu = D_\mu X_\nu.
\eqn\manzana
$$
Notice that we are considering the model presented in [\ag] with only one
holomorphic Killing vector. This is the situation which leads to the
topological quantum field theory constructed in the previous section.

An important remark in the twisting of the topological sigma model leading to 
the action \dos\ is the following. $N=2$ supersymmetric sigma models exist
for flat two-dimensional manifolds. Their formulation on curved manifolds
implies the introduction of $N=2$ supergravity. The twisting is indeed done
on a flat two-dimensional manifold. Once the flat action is obtained one
keeps only one half of the two initial supersymmetries and studies if the
model exist for curved manifolds. It turns out that it exists endowed with
that part of the supersymmetry, a symmetry, $Q$, which is odd and scalar and
often called topological symmetry, and that the resulting action is
\dos. This procedure is standard in any twisting process. One might find,
however, that in order to have invariance under the
topological symmetry, $Q$, it is necessary to add extra terms involving the
curvature to the covariantized twisted action. As we will discuss in the next
subsection this will be the case when considering non-abelian monopoles.

The $Q$-transformations of the fields are easily derived from the $N=2$
supersymmetric transformations in [\ag]. They turn out to be:
$$
\eqalign{
[Q,x^\mu] =& i\chi^\mu, \cr
\{Q,\chi^\mu\} =& -iX^\mu, \cr
\{Q,\rho_\alpha^\mu\} =& \partial_\alpha x^\mu+\epsilon_\alpha{}^\beta
J^\mu{}_\nu\partial_\beta x^\nu - i\Gamma_{\nu\sigma}^\mu
\chi^\nu\rho_\alpha^\sigma,\cr} \eqn\cuatro
$$
where $\epsilon$ is the complex structure induced by $h$ on $\Sigma$.
As it is the case for the $N=2$
supersymmetric transformations in [\ag], this symmetry is realized
on-shell. After using the field equations one finds:
$$
\eqalign{
[Q^2, x^\mu] =& X^\mu, \cr
[Q^2, \chi^\mu] =& \partial_\nu X^\mu \chi^\nu, \cr
[Q^2, \rho_\alpha^\mu] =& \partial_\nu X^\mu \rho_\alpha^\nu. \cr}
\eqn\cinco
$$
From these relations one can read off the action of the central-charge
generator in \orangedos: $Z$ acts as a Lie derivative with respect to the
vector field $X^\mu$. This is exactly the action found for $Q^2$ in the
previous section (see eq. \otrolio). In addition,
it is straightforward to verify that the first two transformations in 
\cuatro\ are the same as the ones generated by $d_{\hat X}$ in subsection 4.1.
In order to compare the transformation for $\rho_\alpha^\mu$ in \cuatro\ to
the one in \brstfibra\ we need first to introduce auxiliary fields to
reformulate the twisted theory off-shell. As shown in [\topmodel,\tmintwo],
this is easily achieved twisting the off-shell version of the $N=2$
supersymmetric theory. In the twisted theory these auxiliary fields, which
will be denoted as $H_\alpha^\mu$, can be understood as a basis on the fibre
${\cal V}_f$. Coming from an off-shell untwisted theory, 
they enter in the twisted
action quadratically. As expected, after adding the topological invariant
term, 
$$
S_2= {1\over 2} \int_\Sigma d^2z \sqrt{h}
 \epsilon^{\alpha\beta} J_{\mu\nu} \partial_\alpha x^\mu
\partial_\beta x^\nu,
\eqn\seis
$$
the action of the off-shell twisted theory can be written in a $Q$-exact form:
$$
\Big\{Q, \int_\Sigma \sqrt{h}\Big[
{1\over 2}h^{\alpha\beta}G_{\mu\nu} \rho_\alpha^\mu(\partial_\beta x^\nu
-{1\over 2}H^\nu_\beta) + i G_{\mu\nu}X^\mu\chi^\nu\Big]\Big\}=
S_1 + S_2 - {1\over 4} \int_\Sigma
\sqrt{h}h^{\alpha\beta}G_{\mu\nu}H^\mu_\alpha H^\nu_\beta,
\eqn\siete
$$
where one has to take into account the
$Q$-transformation of the auxiliary field $H_\alpha^\mu$ and the corresponding
modifications of the  third $Q$-transformation in \cuatro:
$$
\eqalign{
\{Q,\rho_\alpha^\mu\} = & H^\mu_\alpha +
\partial_\alpha x^\mu+\epsilon_\alpha{}^\beta  
J^\mu{}_\nu\partial_\beta x^\nu -
i\Gamma_{\nu\sigma}^\mu\chi^\nu\rho_\alpha^\sigma, \cr
[Q,H^\mu_\alpha] = & -iD_\alpha\chi^\mu -i\epsilon_\alpha{}^\beta  
J^\mu{}_\nu D_\beta\chi^\nu - i\Gamma_{\nu\sigma}^\mu\chi^\nu H_\alpha^\tau -
{1\over 2} R_{\sigma\nu}{}^\mu{}_\tau\chi^\sigma\chi^\nu\rho_\alpha^\tau
+D_\tau X^\mu \rho_\alpha^\tau. }
\eqn\ocho
$$

The auxiliary field $H^\mu_\alpha$ entering \siete\ and \ocho\ is not the
same as the one in \brstfibra\ and \gauge. Notice that in the action
resulting after computing $Q\Psi$ in \gauge\ the auxiliary field does not
enter only quadratically in the action. A linear term is also present. In 
\siete, however, only a term quadratic in $H^\mu_\alpha$ appears. Also the 
transformations \ocho\ and \brstfibra, as well as the gauge fermion in
\siete\ and \gauge, are different. Redefining the auxiliary field 
$H^\mu_\alpha$ as:
$$
\Pi^\mu_\alpha = H^\mu_\alpha +
\partial_\alpha x^\mu+\epsilon_\alpha{}^\beta  
J^\mu{}_\nu\partial_\beta x^\nu -
i\Gamma_{\nu\sigma}^\mu\chi^\nu\rho_\alpha^\sigma,
\eqn\laurita
$$
one finds that:
$$
\eqalign{
\{Q,\rho_\alpha^\mu\} = & \Pi^\mu_\alpha, \cr
[Q,\Pi^\mu_\alpha] = & {\partial}_\tau X^\mu \rho_\alpha^\tau, }
\eqn\laspis
$$
and  the resulting action has the form:
$$
\Big\{Q, \int_\Sigma \sqrt{h}\Big[
h^{\alpha\beta}G_{\mu\nu} \rho_\alpha^\mu(\partial_\beta x^\nu
-{i\over 4}\Gamma_{\sigma\tau}^\nu\chi^\sigma\rho_\beta^\tau-{1\over
4}\Pi^\nu_\beta) + i G_{\mu\nu}X^\mu\chi^\nu\Big]\Big\}.
\eqn\martia
$$
This action differs from the one  that follows after acting with $Q$ on
the gauge fermion \gauge\ in the terms which are originated from 
$ i G_{\mu\nu}X^\mu\chi^\nu$ in \martia. 
These are precisely the terms obtained in \masterminos\ in the previous
section. Thus the twisted theory corresponds to the one obtained from
the equivariant extension of the Mathai-Quillen formalism once the
localization term \masterminos\ is added. Notice that from the point of view
of the equivariant extension of the Mathai-Quillen formalism this additional
term can be introduced with an arbitrary multiplicative constant $t$. 
Since the dependence on
the parameter $u$ of section 2 can be reabsorbed in the vector field $X$, one
has a one-parameter family of actions for a fixed Killing vector $X$.
Since this parameter enters only in a $Q$-exact term one expects that no
dependence on it appears in vacuum expectation values, at least if some 
requirements on compactness are fulfilled. This opens new ways to
compute topological invariants by considering different limits of this
parameter, and the resulting approach corresponds mathematically 
to localization of 
integrals of equivariant forms. The simplest case, the homotopically 
trivial maps from the Riemann surface $\Sigma$ to the target space $M$, was 
explicitly considered in [\llla], and some classical localization results like 
the Poincar\'e-Hopf theorem were rederived in this framework.

As discussed in the previous
section, this topological theory, as the non-extended one,
can be generalized to the case of an almost-hermitean manifold. We will no
describe here this generalization. The existence of this generalization was
first discussed in [\llla] and, as shown in sect. 4.2, it can also be
formulated from an equivariant extension of the Mathai-Quillen formalism.

\section{Non-abelian monopoles}

We will begin recalling  the structure of $N=2$ supersymmetric Yang-Mills
coupled to massive $N=2$ supersymmetric matter fields. The pure Yang-Mills part
is built out of an $N=2$ vector multiplet which contains a vector field
$A_\mu$, a right-handed spinor $\lambda^{i}_{\alpha}$, a left-handed spinor
$\bar\lambda_{i\dot\alpha}$ and a complex scalar $B$.
The twisting of this part of the model leads to Donaldson-Witten
theory [\tqft]. 
$N=2$ supersymmetric matter fields are introduced with the help
of hypermultiplets. A hypermultiplet contains two complex bosonic fields
$q^i$ which  transform as an $SU(2)_I$ isodoublet, two right-handed
spinors, $\psi_{q\alpha}$ and $\psi_{\tilde q\alpha}$, and two left-handed
spinors $\bar\psi_{\tilde q\dot\alpha}$ and $\bar\psi_{q\dot\alpha}$, all
transforming as a scalar under $SU(2)_I$. The twisting of hypermultiplets
coupled to $N=2$ supersymmetric Yang-Mills has been considered in 
[\mart,\alab,\ans,\alabas,\ansdos,\tqcd]. 
Under the twsiting the fields become:
$$
\eqalign{
A_{\alpha\dot\alpha}  & \longrightarrow  \cr
\lambda_\alpha^i &\longrightarrow  \cr
\overline\lambda_{\dot\alpha i} & \longrightarrow \cr
B& \longrightarrow \cr
B^\dagger &\longrightarrow \cr}
\quad
\eqalign{
& A_{\alpha\dot\alpha} , \cr
& \eta  , \;\;\; \chi_{\alpha\beta} , \cr
& \psi_{\alpha\dot\alpha} , \cr
& \lambda , \cr
& \phi , \cr}
\qquad\qquad
\eqalign{
q^i & \longrightarrow \cr
\psi_{q \alpha}  & \longrightarrow \cr
\overline \psi_{\tilde q \dot\alpha} 
& \longrightarrow \cr
q_i^\dagger  & \longrightarrow \cr
\overline \psi_{q \dot\alpha} 
& \longrightarrow \cr
\psi_{\tilde q \alpha} & \longrightarrow \cr}
\quad
\eqalign{    
& M^\alpha , \cr   
& \mu_\alpha , \cr
& v_{\dot\alpha} , \cr
& \overline M_\alpha , \cr   
& \overline v_{\dot\alpha} , \cr
& \overline \mu_\alpha , \cr}
\eqn\paloma
$$
where the field $\chi_{\alpha\beta}$ is symmetric in $\alpha$ and $\beta$ and
therefore it can be regarded as the components of a self-dual two form. 

In order to present the form of the action after the twisting we need to
recall the geometrical
data introduced at the beginning of subsection 4.2. We will be considering a
gauge group $G$ and a principal fibre bundle $P$ on an oriented, closed, spin
four-manifold $X$ endowed with a Riemannian structure given by a metric
$g_{\mu\nu}$.  Then, the field  $A$ represents a $G$-connection  with
associated field strength $F_{\mu\nu}$. For the matter part  
let us consider an associated vector bundle $E$ to
the principal bundle $P$ by means of a representation $R$ of the group $G$.
All the matter fields can be regarded as sections of this vector bundle. The
action which results after the twisting is:
$$
\eqalign{
S_1=&\int_X \sqrt{g}  \Big[\tr\big(
{1\over 4} F_{\mu\nu} F^{\mu\nu}-{i\over \sqrt{2}} \psi^\beta{}_{\dot\alpha}
\nabla^{\alpha\dot\alpha}\chi_{\alpha\beta}
+{i\over 4} \chi^{\alpha\beta} [\phi,\chi_{\alpha\beta}] \cr
&+i\phi \nabla_\mu\nabla^\mu \lambda + {i\over 2}\psi_{\alpha\dot\alpha}
\nabla^{\alpha\dot\alpha}\eta - {1\over 2}\psi_{\alpha\dot\alpha}
[\psi^{\alpha\dot\alpha},\lambda] - {1\over 2} [\phi,\lambda]^2+{i\over 2}
\eta[\phi,\eta] \big) \cr
&+\nabla_\mu \overline M^\alpha \nabla^\mu M_\alpha
+{1\over 4} R \overline M^\alpha M_\alpha
-{1\over 8} \overline M^{(\alpha} T^a M^{\beta )}
\overline M_{(\alpha} T^a M_{\beta )} \cr
&-{i\over 2} (\bar v^{\dot\alpha}\nabla_{\alpha\dot\alpha}\mu^\alpha
-\bar\mu^\alpha\nabla_{\alpha\dot\alpha}v^{\dot\alpha})
-i\overline M^\alpha \{\phi,\lambda\} M_\alpha \cr
&+{1\over \sqrt{2}}(\overline M_\alpha \chi^{\alpha\beta} \mu_\beta
-\bar \mu_\alpha \chi^{\alpha\beta} M_\beta)
-{1\over 2}(\overline M^\alpha \psi_{\alpha\dot\alpha} v^{\dot\alpha}
-\bar v^{\dot\alpha}\psi_{\alpha\dot\alpha} M^\alpha ) \cr
&-{1\over 2}(\bar \mu^\alpha \eta M_\alpha + \overline M^\alpha\eta\mu_\alpha)
+{i\over 4} \bar v^{\dot\alpha} \phi v_{\dot\alpha} - \bar\mu^\alpha \lambda
\mu_\alpha \cr
&+{1\over 4} m^2 \overline M^\alpha M_\alpha + {1\over 4} m \bar \mu^\alpha
\mu_\alpha - {1\over 4} m \bar v^{\dot \alpha}v_{\dot\alpha}
- m \overline M^\alpha \lambda M_\alpha - {i\over 4} m \overline M^\alpha\phi
M_\alpha \Big],\cr}
\eqn\laaccion
$$
where $m$ is a mass parameter.
Notice the presence of a term involving the curvature of the four-manifold
$X$. This term must enter the action in order to preserve invariance
under the topological symmetry $Q$ on curved manifolds. 
Notice also that the matter fields with bars carry a representation
$\overline R$ conjugate to $R$, the one carried by the matter fields
without bars. The $Q$ transformations of the fields are: 
$$
\eqalign{
[Q, A_\mu] = & \psi_\mu, \cr 
\{Q,\psi_\mu \} = & \nabla_\mu \phi, \cr
[Q, \lambda] = & \eta, \cr 
\{Q,\eta\} = & i[\lambda,\phi], \cr
[Q, \phi] = & 0, \cr 
\{Q,\chi^a_{\alpha\beta} \} = & -i\sqrt{2} 
(F^a_{\alpha\beta}+{i\over 2}\overline M_{(\alpha} T^a M_{\beta)}), \cr}
\qquad\qquad
\eqalign{
[Q, M_\alpha] = & \mu_\alpha, \cr 
[Q, \overline M_\alpha] = & \bar \mu_\alpha, \cr 
\{Q,\mu_\alpha \} = & m M_\alpha - i \phi M_\alpha, \cr
\{Q, \bar\mu_\alpha \} = &- m \overline M_\alpha + i \overline M_\alpha \phi,
\cr
\{Q,v_{\dot\alpha}\} = & -2i\nabla_{\alpha\dot\alpha} M^\alpha,\cr
\{Q,\bar v_{\dot\alpha}\} = 
& -2i\nabla_{\alpha\dot\alpha}\overline M^\alpha.\cr}
\eqn\dadi
$$
These transformations close on-shell up to a gauge transformation whose gauge
parameter is the scalar field $\phi$ and up to a central charge transformation
of the type presented in \orange\ whose parameter is proportional to the mass
of the field involved:
$$
\eqalign{
[Q^2, A_\mu] = & \nabla_\mu \phi, \cr 
\{Q^2,\psi_\mu \} = & i[\psi_\mu,\phi], \cr
[Q^2, \lambda] = & i[\lambda,\phi], \cr 
\{Q^2,\eta\} = & i[\eta,\phi], \cr
[Q^2, \phi] = & 0, \cr 
\{Q^2,\chi_{\alpha\beta} \} = & i[\chi_{\alpha\beta},\phi], \cr}
\qquad\qquad
\eqalign{
[Q^2, M_\alpha] = & m M_\alpha - i \phi M_\alpha, \cr 
[Q^2, \overline M_\alpha] = & -m \overline M_\alpha + i \overline M_\alpha
\phi, \cr 
\{Q^2,\mu_\alpha \} = & m \mu_\alpha - i \phi \mu_\alpha, \cr
\{Q^2, \bar\mu_\alpha \} = &-m \bar\mu_\alpha + i \bar\mu_\alpha \phi, \cr
\{Q^2,v_{\dot\alpha}\} = & m v_{\dot\alpha} - i \phi v_{\dot\alpha},\cr
\{Q^2,\bar v_{\dot\alpha}\} = &- m \bar v_{\dot\alpha} 
+ i \bar v_{\dot\alpha} \phi.\cr}
\eqn\dadidos
$$
Notice that for the last transformation in the first set and for the last
two in the second set we have made use of the field equations. 
The central charge acts trivially on the pure Yang-Mills fields or
Donaldson-Witten fields but non-trivially on the matter fields. As it will
become clear in the forthcoming discussion this symmetry is precisely the
$U(1)$ symmetry entering the equivariant extension carried out in subsection
4.2. Notice also that the transformations in \dadi\ are the ones generated
by $d_{X_{\cal P}}$ in that subsection, with $im$ playing the role of the
parameter $u$, except for the  fields $\chi_{\alpha\beta}$, $v_{\dot\alpha}$
and $\bar v_{\dot\alpha}$. In fact, the mass terms in \laaccion\ are precisely 
\masillas\ and \masotta. The terms coming from the $d_{X_{\cal P}}$-exact term 
can have an arbitrary multiplicative parameter $t$, \ie, they enter in 
the exponential of \lagrangiano\ as $t Q \omega_{X_{\cal P}}$. This parameter 
must be $t=-im/4$ in order to recover the twisted theory (notice that
the exponential of \lagrangiano\ has to be compared to minus the action of
the twisted theory).

Our next goal is, as in the case of topological sigma models, to construct an
off-shell version of the twisted model. There are two possible ways to
build an off-shell version. One could consist of considering
off-shell versions of $N=2$ supersymmetry. This have been analyzed in
[\mart,\alab,\alabas] showing that it does not lead to a formulation whose
action is $Q$-exact. As shown for first time in [\mart] one needs to
introduce an auxiliary field different than the one originated
 from the off-shell supersymmetric theory in order to have an off-shell
formulation with a $Q$-exact action. This is precisely the same conclusion
that is achieved considering a second way to construct an off-shelf
formulation. In this alternative approach the steps to be followed are the
same ones as in the case of the topological sigma models: introduce auxiliary
fields $K_{\alpha\beta}$,  $k_{\dot\alpha}$ and $\bar k_{\dot\alpha}$ 
in the transformations of
$\chi_{\alpha\beta}$, $v_{\dot\alpha}$ and $\bar v_{\dot\alpha}$ respectively,
and define the transformations of these fields in such a way that $Q^2$ on 
$\chi_{\mu\nu}$, $v_{\dot\alpha}$ and $\bar v_{\dot\alpha}$ closes without
making use of the field equations. Following this approach one finds:
$$
\eqalign{
\{ Q , \chi_{\alpha\beta}^a \} = & K^a_{\alpha\beta}
-i\sqrt{2}( F^a_{\alpha\beta}+{i\over 2}\overline M_{(\alpha} 
T^a M_{\beta)}), \cr
\{ Q , v_{\dot\alpha} \} = & k_{\dot\alpha}
-2i \nabla_{\alpha\dot\alpha} M^\alpha, \cr
\{ Q , \bar v_{\dot\alpha} \} = & \bar k_{\dot\alpha}
-2i \nabla_{\alpha\dot\alpha} \overline M^\alpha, \cr
[ Q , K_{\alpha\beta}^a ] = &
i[\chi_{\alpha\beta}, \phi]^a -{i\over\sqrt{2}} [\nabla_{(\alpha\dot\beta}
\psi_{\beta)}{}^{\dot\beta}]^a + {1\over \sqrt{2}}(\bar\mu_{(\alpha}
T^a M_{\beta)} + \overline M_{(\alpha} T^a \mu_{\beta)}),\cr
[ Q , k_{\dot\alpha} ] = & m v_{\dot\alpha} - i\phi v_{\dot\alpha} - 2
\psi_{\alpha\dot\alpha}M^\alpha + 2i \nabla_{\alpha\dot\alpha} \mu^\alpha,\cr
[ Q , \bar k_{\dot\alpha} ] = & 
- m v_{\dot\alpha} + i\phi v_{\dot\alpha} - 2
\psi_{\alpha\dot\alpha}\overline M^\alpha + 2i \nabla_{\alpha\dot\alpha}\bar
\mu^\alpha. \cr} 
\eqn\pendulo
$$
The non-trivial check now is to verify that $Q^2$ on the auxiliary fields
closes properly. One easily finds that this is indeed the case:
$$\eqalign{
[ Q^2 , K_{\alpha\beta} ] = & i[K_{\alpha\beta},\phi],\cr
[ Q^2 , k_{\dot\alpha} ] = & m k_{\dot\alpha} - i\phi k_{\dot\alpha},\cr
[ Q^2 , \bar k_{\dot\alpha} ] = & - m k_{\dot\alpha} + i\bar
k_{\dot\alpha}\phi.\cr} 
\eqn\pendulodos
$$
It is important to remark that these relations imply that $Q$ closes
off-shell. Our next task is to show that $S_1$ is equivalent to a $Q$-exact
action.

After adding the topological invariant term involving the Chern class,
$$
S_2 = {1\over 4} \int_X F\wedge F,
$$
one finds that the off-shell twisted action of the model can be written as a
$Q$-exact term:
$$
\{Q, \Lambda_0 \} =   S_1+ S_2 + {1\over
4}\int_X \sqrt{g} \big(K^{\alpha\beta} K_{\alpha\beta} +\bar k^{\dot\alpha}
k_{\dot\alpha}\big),
\eqn\burbuuno 
$$
where,
$$
\eqalign{
\Lambda_0 = & \int_X \sqrt{g} \Big[ {1\over
4}\chi^a_{\alpha\beta}
\big(i\sqrt{2}(F^a_{\alpha\beta}
+{i\over 2} \overline M_{(\alpha}T^a M_{\beta)})+
K^a_{\alpha\beta} \big)\cr
&+{1\over 8} \bar v^{\dot\alpha} ( 2 i
\nabla_{\alpha\dot\alpha} M^\alpha +k_{\dot\alpha}) 
-{1\over 8}( 2 i
\nabla_{\alpha\dot\alpha} \overline M^\alpha +\bar k_{\dot\alpha})  
v^{\dot\alpha} \cr
&+\tr\big(i\lambda\nabla_{\alpha\dot\alpha}\psi^{\alpha\dot\alpha}-{i\over 2}
\eta [\phi,\lambda]\big) -{1\over 2}(\bar\mu^\alpha\lambda M_\alpha
-\overline M^\alpha\lambda \mu_\alpha) - {1\over 8} m (\bar\mu^\alpha M_\alpha
-\overline M^\alpha \mu_\alpha ) \Big]\cr}
\eqn\burbu
$$
The auxiliary field entering \burbu\ is not the same  as the one entering
\alsacia. Again, the auxiliary fields $K_{\alpha\beta}$, $k_{\dot\alpha}$
and $\bar k_{\dot\alpha}$
in \burbu\ appear only quadratically in the action, contrary to the way they
appear in the Mathai-Quillen formalism. The relation between these two sets
of fields can be easily read comparing \fermiones\ and \burbu, or \alsacia\
and \pendulo. Redefining the auxiliary fields as
$$
\eqalign{
H_{\alpha\beta}^a = & K^a_{\alpha\beta}
-i\sqrt{2}( F^a_{\alpha\beta}+{i\over 2}\overline M_{(\alpha} 
T^a M_{\beta)}), \cr
h_{\dot\alpha} = & k_{\dot\alpha}
-2i \nabla_{\alpha\dot\alpha} M^\alpha, \cr
\bar h_{\dot\alpha} = & \bar k_{\dot\alpha}
-2i \nabla_{\alpha\dot\alpha} \overline M^\alpha, \cr}
\eqn\jas
$$
one finds that,
$$
\eqalign{
[ Q , H_{\alpha\beta} ] = & i[H_{\alpha\beta},\phi],\cr
[ Q , h_{\dot\alpha} ] = & m v_{\dot\alpha} - i\phi v_{\dot\alpha},\cr
[ Q , \bar h_{\dot\alpha} ] = 
& - m v_{\dot\alpha} + i \bar v_{\dot\alpha}\phi ,\cr}
\eqn\muelle
$$
and the resulting action takes the form:
$$
\{Q, \Lambda \}
\eqn\osci
$$
where,
$$
\eqalign{
\Lambda = & \int_X \sqrt{g} \Big[ {1\over
4}\chi^a_{\alpha\beta}\big(i2\sqrt{2}(F^a_{\alpha\beta}
+{i\over 2} \overline M_{(\alpha}T^a M_{\beta)})
+ H^a_{\alpha\beta}\big)\cr
&+{1\over 8}\bar v^{\dot\alpha} (4i\nabla_{\alpha\dot\alpha} 
M^\alpha +h_{\dot\alpha})-{1\over 8}(4i \nabla_{\alpha\dot\alpha} 
\overline M^\alpha +
\bar h_{\dot\alpha}) v^{\dot\alpha}
\cr
&+\tr\big(i\lambda\nabla_{\alpha\dot\alpha}\psi^{\alpha\dot\alpha}-{i\over 2}
\eta [\phi,\lambda]\big) -{1\over 2}(\bar\mu^\alpha\lambda M_\alpha
-\overline M^\alpha\lambda \mu_\alpha) - {1\over 8} m (\bar\mu^\alpha M_\alpha
-\overline M^\alpha \mu_\alpha ) \Big]\cr}
\eqn\burbutres
$$

The action \osci\ differs from the one that follows after acting with $Q$ on
the gauge fermions \fermiones\ in the terms which are originated from
$-\tr({i\over 2} \eta [\phi,\lambda])$ and from
$-{1\over 8} m (\bar\mu^\alpha M_\alpha -\overline M^\alpha \mu_\alpha )$.
The absence of a term like the first of these two in the Mathai-Quillen
formalism is a well known fact. It is believed that its presence does not
play any important role towards the computation of topological invariants.
Respect to the second term, it turns out that it has the same origin as
the extra term appearing in the case of topological sigma models with
potentials. This term is precisely the localization term discussed in 
\masotta\ and from a geometrical point of view it has the same origin as
\masterminos. Again, this term can be introduced with an arbitrary constant
providing a model in which an additional parameter can be introduced.
As in the case of topological sigma models one would expect that the vacuum
expectation values of the observables of the theory are independent of this
parameter, and therefore that one can localize this computation 
to the fixed points of the $U(1)$ symmetry, as it has been
 argued in [\pids] from a 
different point of view.

\endpage

\chapter{Conclusions}

In this paper we have obtained equivariant extensions of the Thom form 
with respect to 
a vector field action, in the framework of the Mathai-Quillen formalism. This
construction can be regarded as a generalization of the equivariant curvature 
constructions considered by Atiyah and Bott 
and Berline and Vergne. Furthermore, we 
have shown that this equivariant extension corresponds to the 
topological action of twisted $N=2$ supersymmetric theories with a 
central charge. The formalism we have introduced gives 
a unified framework to 
understand the topological structure of this kind 
of models. The appearance of 
potential or mass terms in twisted $N=2$ theories has been 
sometimes misleading, because one can think that these additional 
terms spoil 
the topological invariance of the theory. As we have 
shown, these models have 
a very simple topological structure in terms of 
equivariant cohomology with 
respect to a vector field action, and of the corresponding 
equivariant extension 
of the Mathai-Quillen form. We also have analyzed in detail two explicit 
realizations of this formalism: topological sigma models 
with a Killing, almost 
complex action on an almost hermitean target space, 
and topological Yang-Mills 
theory coupled to twisted massive hypermultiplets. 

There are other moduli problems, as the Hitchin 
equations on Riemann surfaces, with a $U(1)$ symmetry or a 
vector field action 
similar to the ones considered in this paper. It would be 
interesting to study 
their Mathai-Quillen formulation and its equivariant extension, and to relate 
them to twisted supersymmetric theories. But perhaps the
 most interesting extension of our work is to implement 
the localization theorems of equivariant cohomology in this framework. It has 
been shown in [\ab, \berlin] that the integral of a closed equivariant 
differential form can be always restricted to the fixed points of the 
corresponding $U(1)$ or vector field action. This can be used 
to relate, for 
instance, characteristic numbers to quantities associated to this zero locus. 
The topological invariants associated to topological sigma models and 
non-abelian monopoles on four-manifolds can be understood as integrals of 
differential forms on the corresponding moduli spaces. In the first case we 
get the Gromov invariants, and in the second case a generalization of the 
Donaldson 
invariants for four-manifolds. If we consider 
the equivariant extension of 
these models, we could compute the topological 
invariants in terms of adequate 
restrictions of the equivariant integration 
to the zero locus of the corresponding abelian symmetry. In fact, it has been 
argued in [\pids] that localization techniques can 
provide a explicit link 
between the Donaldson and the Seiberg-Witten invariants, 
because their 
moduli spaces are precisely the fixed points of the abelian $U(1)$ symmetry 
considered in \circulos, acting on the moduli space of $SU(2)$ 
monopoles. Perhaps the techniques of equivariant integration, 
applied to the equivariant differential forms considered in this paper,
can give an explicit proof of this link. However, a key point when one tries 
to apply localization techniques is the compactness of the moduli spaces. The 
vector field action can have fixed points on the compactification divisors 
which give crucial contributions to the equivariant integration. This situation 
arises in both the topological sigma model and the non-abelian monopoles on 
four manifolds. It can be easily seen that, without taking into account the 
compactification of the moduli space, one doesn't obtain sensible results for 
the quantum cohomology rings or the polynomial invariants of four-dimensional 
manifolds.

In our four-dimensional example we have seen that the equivariant extension 
of the non-abelian monopole theory corresponds to the twisted $N=2$ Yang-Mills 
theory coupled to massive hypermultiplets. It would be very interesting to 
use the exact solution of the physical theory given in [\ws] to obtain 
the topological correlators of the twisted theory, as it has been 
done in [\wjmp, \wv, \mfm, \lmdos]. 
It seems that the duality structure of $N=2$ and 
$N=4$ gauge 
theories ``knows" about the compactification of the moduli space of their 
twisted 
counterparts, and therefore the physical approach would shed new light on the 
localization problem.

\vskip1cm

\ack
We would like to thank C. Lozano for important remarks concerning the
twisting of $N=2$ supersymmetric theories. M.M. would like to thank V.
Pidstrigach for useful correspondence, J.A. Oubi\~na for many  useful
discussions and a careful reading of the manuscript, and the Theory 
Division at CERN for its hospitality. This work was supported
in part by DGICYT under grant PB93-0344 and  by CICYT under grant AEN94-0928.

\endpage

\refout
\end